\documentclass[12pt]{article}
\newcommand{\eps}{\epsilon}
\newcommand{\cR}{\mathcal{R}} \newcommand{\cJ}{\mathcal{J}} \newcommand{\cP}{\mathcal{P}}
\newcommand{\cD}{\mathcal{D}} 
\newcommand{\cE}{\mathcal{E}}
\newcommand{\bD}{\mathbb{D}} \newcommand{\bA}{\mathbb{A}}
\newcommand{\bE}{\mathbb{E}}
\newcommand{\SN}{\mathrm{su}_{N;}{}}
\newcommand{\Ss}{\mathrm{su}_{N;}^2{}}
\newcommand{\su}{\mathrm{su}_{3;}^2{}}
\usepackage{amssymb}      

\newtheorem{theorem}{Theorem}[section]

\newtheorem{lemma}{Lemma}[section]
\newtheorem{corollary}{Corollary}[section]

\newcommand{\ii}{\mathrm{i}}

\begin{document}



\centerline{{\LARGE The $W_N$ minimal model classification}}

\bigskip\centerline{Elaine Beltaos}

\centerline{{\it Mathematics Department, Grant MacEwan University}}

\centerline{{\it 10700 - 104 Ave, Edmonton, AB CANADA T5J 4S2}}

\bigskip\centerline{Terry Gannon}\medskip

\centerline{{\it Mathematics Department, University of Alberta}}

\centerline{{\it Edmonton AB CANADA T6G 2G1}}\bigskip

\centerline{ BeltaosE@macewan.ca}

\centerline{tgannon@math.ualberta.ca}\bigskip

\noindent{\bf Abstract.}
We first rigourously establish, for any $N\ge 2$, that the  toroidal modular
invariant partition functions for the (not necessarily unitary)
$W_N(p,q)$ minimal models \textit{biject} onto a well-defined subset of those of the
 SU$(N)\times\,$SU$(N)$ Wess-Zumino-Witten theories at level $(p-N,q-N)$. This permits
 considerable simplifications to the
 proof of the Cappelli-Itzykson-Zuber classification of Virasoro minimal models.
More important, we obtain from this the complete classification of  all modular invariants
for the $W_3(p,q)$ minimal models. All should be realised by rational conformal field theories.
Previously, only those for the unitary models, i.e. $W_3(p,p+1)$, were classified.
For all $N$ our correspondence yields for free an extensive  list of $W_N(p,q)$ modular invariants.
The $W_3$ modular invariants, like the Virasoro minimal models, all factorise into SU$(3)$
modular invariants, but this fails in general for larger $N$. We also classify the SU$(3)\times\,$SU(3)
modular invariants, and find there a new infinite series of exceptionals.
\bigskip

\section{Introduction}\label{sIntro}
Each chiral half of a rational conformal field theory (RCFT) is controlled by
 a chiral algebra (rational vertex operator algebra).
These associate to each surface a finite-dimensional
space of conformal blocks, possessing appropriate conditions of analyticity, factorisation,
monodromy etc.

The quantities in the full RCFT of greatest interest are certain sesquilinear  combinations of
conformal blocks called correlation functions. A beautiful theory (see e.g. \cite{RFFS}),
related at least in broad strokes to the subfactor approach to RCFT (see e.g.  \cite{BE}),
 finds these sesquilinear combinations starting from
a special symmetric Frobenius algebra of the category of modules of the chiral algebra.
This theory
regards the 1-loop open string partition function (\textit{nim-rep}) as more fundamental than the
1-loop closed string partition function (\textit{modular invariant}). However, the latter is far more tightly constrained
and within this framework the classification of RCFT associated to a given chiral algebra would (except for the easiest examples) proceed
first by classifying the modular invariants, then the corresponding nim-reps, and
then from that the possible Frobenius algebras. For example,  if the modular invariant
is a permutation matrix, then the corresponding Frobenius algebra is Azumaya. More precisely, the baby case, namely SU(2) Wess-Zumino-Witten (WZW),
is the exception as its nim-reps correspond to graphs with largest eigenvalue $<2$
and so are easily classified. But already SU(3) WZW nim-reps seem hopeless
 to classify  and include infinitely many `nonphysical' nim-reps (the first at level 3, given by
 quantum-dimension). `Nonphysical' here means nim-reps not corresponding to a modular invariant.
The point is that, even in this categorical framework, the starting point for the RCFT classification
in practice would be  the modular invariant classification.

The Virasoro minimal models  constitute perhaps the best known RCFTs. They
belong to a sequence of reasonably
accessible rational chiral algebras: the $W_N(p,q)$ minimal models for $N\ge 2$,
 generated by fields of conformal weight $2,3,\ldots,
N$. Here  $p,q$ are \textit{coprime} integers, i.e. gcd$(p, q)
= 1$. $W_2$ recovers the Virasoro algebra, while $W_3$ was introduced by A. B. Zamolodchikov
\cite{Zam}, and the generalisation to higher $N$ is due to Fateev-Lukyanov \cite{FL}  (see also
\cite{BS1,BS2}). The minimal models $W_N(p,q)$ are related
for instance to fractional level admissible modules of affine $A_{N-1}$
(see e.g. \cite{FKW,MW}). For instance, $W_N(p,q)$ is realised by the  Goddard-Kent-Olive
diagonal coset $\mathrm{su}(N)_k\oplus\mathrm{su}(N)_1/\mathrm{su}(N)_{k+1}$ where the
level $k$ is $p/|q-p|-N$.

The modules of a rational chiral algebra naturally inherit a Hermitian inner product.
 An RCFT is called \textit{unitary} if this inner product is positive-definite.
For example the WZW models (i.e. the RCFT associated to compact groups and affine algebras at integral level)
are unitary,  while the $W_N$ minimal models are  typically non-unitary.
Most attention in the literature has focussed on  {unitary}
RCFT, but this seems unwarranted. For example, in the statistical mechanical realisation of RCFTs
unitarity has no physical significance --- for example, 2-dimensional ferromagnets at
criticality are described by the non-unitary  Yang--Lee model, $W_2(2,5)$. 
Moreover, in string theory
the matter CFT is coupled to the non-unitary (super-)ghost CFT;
what must be unitary is the corresponding BRST cohomology, but the relationship between
the (non)unitarity of  the matter CFT and that of BRST cohomology isn't obvious.
 Also, very little marks the structural distinction between unitary and non-unitary vertex operator algebras. 

However, from the point of view of modular invariant classifications, non-unitarity \textit{does}
introduce significant additional challenges, ultimately because the vacuum is not the
primary of minimal conformal weight. Addressing a large nontrivial
class of these is the main motivation for this paper. (The only previous non-unitary classifications 
were $W_2(p,q)$ and SU$(2)$ at fractional level, and these could avoid these aforementioned 
serious challenges through a technicality unavailable for $N>2$, as explained in Section 3.3.)

In particular, non-unitary $W_N(p,q)$ looks more akin  to  WZW at \textit{fractional}  than
\textit{integral} level (recall the coset description given above). Modular invariant classifications for
fractional level WZW are notoriously wild (see \cite{Lu} for the SU(2) story). Is the tameness
of the `A-D-E'  classification  \cite{CIZb,CIZ} for $W_2$ a low rank accident?

More precisely, the $W_N(p,q)$ \textit{modular data} (i.e. the matrices $S$ and $T$ defining
the modular group representation) looks like that of SU$(N)\times\mathrm{SU}(N)$
WZW modular data at fractional heights  $p/q$ and $q/p$ (height$\,=N+\,$level). Applying to
this modular data the {\it Galois shuffle} of 
\cite{Nonunitary}, we associate to each $W_N(p,q)$ minimal model
a unique  SU$(N)\times\mathrm{SU}(N)$  modular invariant at \textit{integral} heights $p,q$.
For $N=2$ and $N=3$, the SU$(N)\times\mathrm{SU}(N)$ modular invariants corresponding to $W_N$ minimal
models can always be expressed in terms of those
of SU$(N)$ together with  simple-current modular invariants (though the {proof}  for
$N=3$ is difficult, as we see in this paper). Even if this pattern were to continue for higher $N$,
 the complete list of SU($N$) WZW
modular invariants is not known for $N>3$, making the $W_N$ classification
for higher $N$ out of reach for now.

The $W_2(p,q)$ minimal model classification (both unitary and non-unitary) is due to
  Cappelli--Itzykson--Zuber \cite{CIZb}. Our $W_N\hookrightarrow$ WZW relation permits 
significant simplifications to  this proof. The \textit{unitary} $W_3$ minimal models, i.e.
  $W_3(p,p+1)$, were found in \cite{coset} using the coset realisation $\mathrm{su}(3)_{p-3}\oplus
  \mathrm{su}(3)_1/\mathrm{su}(3)_{p-2}$.  In this paper, we use the $W_N\hookrightarrow$ WZW
  relation to obtain the full $W_3(p,q)$ classification (i.e. non-unitary as well as unitary).
 This project (obtaining the $W_3$ minimal
model classification via the SU$(3)\times\mathrm{SU}(3)$ one) constituted parts of the first author's
theses \cite{ThesisM,ThesisP}. As was the case for $W_2$, \textit{all of the $W_3$ modular invariants
(but only about half of the $\mathrm{SU}(3)\times\mathrm{SU}(3)$ ones)  come in factorised form.} All  $W_3$
modular invariants have a corresponding nim-rep and we expect will arise as the partition function
of a healthy RCFT.

This is not true for  SU$(3)\times\mathrm{SU}(3)$. In particular we find a new infinite series of
 SU$(3)\times\mathrm{SU}(3)$ exceptional modular invariants, the first at height $(12,5)$:
 \begin{eqnarray}
&|\chi_{[11](11)}+\chi_{[55](11)}|^2+|\chi_{[11](22)}+\chi_{[55](22)}|^2+|\chi_{[33](31)}+\chi_{[33](13)}|^2
 +|\chi_{[33](12)}+\chi_{[33](21)}|^2\nonumber&\\
&+\,\chi_{[33](11)}(\chi_{[11](31)}+\chi_{[55](31)}+\chi_{[11](13)}+\chi_{[55](13)})^*+
c.c.\nonumber&\\
&+\,\chi_{[33](22)}(\chi_{[11](21)}+\chi_{[55](21)}+\chi_{[11](12)}+\chi_{[55](12)})^*+
c.c.\label{exc2}\ ,&
\end{eqnarray}
where we use the shorthand $\chi_{[ab](cd)}$ for the affine $A_2\oplus A_2$ character
combination
$$\chi_{(12-a-b;a-1,b-1),(5-c-d;c-1,d-1)}+\chi_{(b-1;12-a-b,a-1),(5-c-d;c-1,d-1)}+\chi_{(a-1;b-1,12-a-b),(5-c-d;c-1,d-1)}\,.$$
None of these new exceptionals can be partition functions of an RCFT. The reason is that any
such partition function must be a twist of one of extension type \cite{MS}, i.e. a sum of squares,
and no such `extension type' modular invariant exists for these exceptionals.

An interesting extension of the present work would be  to  extend our $W$-algebra 
$\hookrightarrow$ WZW relation to the
$W$-algebras corresponding to any simple Lie algebra. Explicit classifications
won't be possible for now, except for small level.

We begin with a review of modular invariants  (Section \ref{sModInv}) and  SU($N$)
 modular data  (Section 2.2). Section 3 is the heart of the paper. It reduces the $W_N(p,q)$
minimal model classification to that of SU$(N)\times\mathrm{SU}(N)$ at coprime heights,
and states the $W_3$ and SU(3)$\times$SU(3) classifications. Sections 4 and 5
prove the SU(3)$\times$SU(3) (and hence $W_3$) modular invariant classification. The main results of this paper are
Theorem 3.3 (giving the $W_3$ minimal model classification) and
Theorem \ref{main} (explicitly relating $W_N$  to SU$(N)\times$SU$(N)$).
Also of independent interest should be Theorem 3.2
(giving the SU(3)$\times$SU(3) classification) and Section 3.3 (sketching a new proof for the Virasoro minimal models).

\section{Background}      \setcounter{equation}{0}

For a review of much of the basics of RCFT, see
e.g.\ \cite{DMS}, and for necessary aspects of modular data and
modular invariants, see \cite{moddata}. The
subtleties occurring in non-unitary theories are discussed in
\cite{Nonunitary}. The standard references on $W$-algebras are the
review \cite{BS1} and the reprint volume \cite{BS2}.

\subsection{Modular invariants} \label{sModInv}

The spectrum of an RCFT can be read off from its 1-loop torus
partition function:
\begin{equation}\label{intro:part}
\mathcal{Z} = \sum_{\lambda,\mu\in \Phi}
M_{\lambda\mu}\chi_\lambda\chi_\mu^*\ , \end{equation}
where $\lambda,\mu$ run over the (finitely many) chiral
primaries, parametrising the irreducible modules of
the chiral algebra  of the theory. Let 0 denote the vacuum sector. The
functions $\chi_\lambda$ are conformal blocks for the
torus, and  characters of those modules. Throughout this
paper, $*$ denotes complex conjugation.  The
$\chi_\lambda$ realise a unitary representation of the modular
group SL$_2(\mathbb{Z})$, via
\begin{equation}\label{modularity}
\chi_\lambda\left(\frac{-1}{\tau}\right) = \sum_{\nu\in \Phi}
S_{\lambda\nu}\chi_\nu (\tau)\ ,\qquad \chi_\lambda(\tau+1) =
\sum_{\nu\in\Phi} T_{\lambda\nu}\chi_\nu (\tau)\ . \end{equation}
 The partition function (\ref{intro:part}) must be invariant
under this action of SL$_2(\mathbb{Z})$.

The coefficient matrix $M$ in (\ref{intro:part}) is called a {\it
modular invariant}. More formally, we call $M$ a \emph{modular
invariant} if the following three conditions hold:
\begin{eqnarray}
M_{00} &=& 1 \ \ \qquad\qquad\qquad(\mbox{\emph{uniqueness of
vacuum}}),\label{intro:vac}\\
M_{\lambda\mu}& \in& \mathbb{Z}_{\geq 0} \ \forall \lambda, \mu
\in
\Phi \qquad(\mbox{\emph{integrality and nonnegativity}}),\label{intro:pos}\\
SM&=&MS, \ TM=MT \ \ (\mbox{\emph{modular
invariance}}).\label{intro:modinv}
\end{eqnarray}

The matrices $S,T$ in (\ref{modularity}) are called modular data.
$T$ is a diagonal matrix, with entries
$T_{\lambda\lambda}=\exp[2\pi \ii\,(h_\lambda-\frac{c}{24})]$ where
$h_\lambda$ is the conformal weight and $c$ is the central
charge. The matrix $S$ is more subtle but more important. For
instance the well-known Verlinde's formula expresses the fusion
coefficients in terms of $S$. Let us focus on $S$.

$S$ is a symmetric unitary matrix, $S_{\lambda\mu}=S_{\mu\lambda}$, whose square $S^2$ is an
order-2 permutation matrix called {\it charge-conjugation} $C$:
\begin{eqnarray}
S_{C\lambda, \mu}& =& S_{\lambda, C\mu} = S^{*}_{\lambda\mu}\ ,\label{SSS'}\\
T_{C\lambda, C\mu}& =& T_{\lambda\mu}\ .\label{TTT'}
\end{eqnarray}
 Note that  if $M$ is a modular invariant, so will be the matrix
product $CM=MC$. More generally, if $M'$ is a \textit{permutation invariant}
(see (\ref{perminv}) below), then the products $MM'$ and $M'M$ will also be modular
invariants (though not necessarily equal).

{\it Simple-currents} are permutations $J$ of the primaries $\lambda\in \Phi$, such that
\begin{equation}\label{SSSj}
S_{J\lambda, \mu} = \exp[{2\pi \ii}\,{Q_J}(\mu)]S_{\lambda\mu}\ ,
\end{equation}
for some rational numbers $Q_J(\mu)$.  The primary $J0\in\Phi$ is
also called a simple-current.

The \textit{Galois symmetry} is both the most powerful and the most exotic.
All entries $S_{\lambda\mu}$ lie in the cyclotomic field $\mathbb{Q}[\xi_L]$ for some
root of unity $\xi_L :=\exp[2\pi \ii/L]$.   The Galois
automorphisms $\sigma_\ell$ are parametrised by integers $\ell\in\mathbb{Z}_L^\times$
coprime to $L$, defined mod $L$. For any such $\ell$,  there is a permutation of $\Phi$, also labelled
$\sigma_\ell$, and a choice of signs
$\epsilon_\ell(\lambda)\in \{\pm 1\}$ such that
\begin{equation}\label{GalCond}
\sigma_\ell(S_{\lambda\mu}) =
\epsilon_\ell(\lambda)\,S_{\sigma_\ell\lambda, \mu} =
\epsilon_\ell(\mu)\, S_{\lambda, \sigma_\ell\mu}\ .
\end{equation}
For example, $L=4Nn$ works for $SU(N)$ at level $n$, and $L=4Npq$ works for
$W_N(p,q)$.

The easiest examples of modular invariants are $M=I$
and $M=C$. Another generic source are simple-currents. Let $J$ be of order $d$, where
the conformal weight $h_{J0}$ lies in $\frac{1}{d}\mathbb{Z}$
(this is automatic if $d$ is odd). Then there is a modular
invariant associated to it by
  \begin{equation}\label{scinv}
M[J]_{\lambda\mu}= \sum_{1\le j\le d}\delta^{{\mathbb{Z}}}
\left(Q_J(\lambda)-{j\,h_{J0}}\right)\, \delta_{\mu\,
,J^{j}\lambda}\ ,
\end{equation}
 where $\delta^{{\mathbb{Z}}}(x)$ equals 1 or 0 depending on whether or
not $x$ is integral.

The condition that $M$ commutes with the diagonal matrix $T$ is the selection rule
\begin{equation}\label{TM}
M_{\lambda\mu}\ne 0\ \Rightarrow\ T_{\lambda\lambda}=T_{\mu\mu}\ .
\end{equation}
The symmetries of $S$ become symmetries of
$M$. For example, charge-conjugation obeys
\begin{equation}\label{CM}
M_{C\lambda,C\mu}=M_{\lambda\mu}\ .
\end{equation}
More generally, the Galois symmetry (\ref{GalCond}), $S$-invariance
(\ref{intro:modinv}), and integrality (\ref{intro:pos}) together yield
\begin{equation}\label{GalCondDer}
M_{\lambda\mu} =\epsilon_\ell(\lambda)\,\epsilon_\ell(\mu)\,
M_{\sigma_\ell\lambda, \sigma_\ell\mu}\ .
\end{equation}
If $M_{\lambda\mu} \neq 0$, then (\ref{GalCondDer})
and nonnegativity (\ref{intro:pos}) give the
 {\it parity rule}
\begin{equation}\label{parity}
M_{\lambda\mu} \neq 0 \Rightarrow \epsilon_\ell (\lambda) =
\epsilon_\ell (\mu)\ .
\end{equation}

By the \textit{minimal primary} $o$ we mean the primary $\lambda\in\Phi$ with minimal
conformal weight $h_\lambda$. Equivalently, $o$ is the unique primary obeying
\begin{equation}\label{strict}
S_{\lambda o} \geq S_{0o} > 0\ .
\end{equation}
Equality in (\ref{strict}) will hold iff $\lambda$ is a simple-current.
When $o\ne 0$ (i.e. when the vacuum column of $S$ is not strictly positive), we call the modular
data \textit{non-unitary}. The modular data of WZW models all have $o=0$, i.e. are \textit{unitary},
but most $W_N$ minimal models are non-unitary. (To our knowledge it is
not yet known for every RCFT that there must be a \textit{unique} primary with minimal
conformal weight, but this is true for all $W_N(p,q)$  and all unitary theories.)

The ratios $\cD\lambda:=S_{\lambda o}/S_{0o}$ are called \textit{quantum-dimensions}. Then
 (\ref{strict}) says $\cD\lambda\ge 1$ with equality iff $\lambda$ is a
simple-current. Moreover, $\cD\lambda=\cD\mu$
if (but not in general iff) $\mu=C^aJ\lambda$ for some $a$ and some simple-current $J$.

\begin{lemma}\label{lemma*}  \cite{Revisited} Let $M$ be a modular invariant and $J,J'$ be simple-currents.
Suppose $o=0$.
\begin{itemize}

\item[(a)] If $M_{J0,J'0}\ne 0$ then $M_{J\lambda,J'\mu}=M_{\lambda\mu}$ for all $\lambda,\mu
\in\Phi$. If moreover $M_{\lambda\mu}\ne 0$, then $Q_J(\lambda)\equiv Q_{J'}(\mu)$ (mod 1).

\item[(b)]  For each $\lambda\in\Phi$, define $s_L(\lambda) = \sum_{\mu}
M_{\mu 0}S_{\lambda\mu}$. Then $s_L(\lambda)
 \geq 0$, and $s_L(\lambda) > 0$ iff some $\mu\in\Phi$ has $M_{\lambda\mu}
\ne 0$ (similarly for $s_R(\lambda) = \sum_{\mu} M_{0 \mu}S_{\lambda\mu}$).
\end{itemize}\end{lemma}

This lemma is crucial to the `modern' approach (see Steps 1-3 below) to modular invariant classifications, but fails in general for non-unitary
modular data. Because of this,  non-unitary  modular invariant classifications can
look very different (see \cite{Lu,Nonunitary} for  dramatic
examples) and will in general require new arguments.

Fortunately, in many  non-unitary RCFTs,
the minimal primary and the vacuum are related in a definite way
called the Galois shuffle \cite{Nonunitary}. When this
holds, $M_{oo}=1$ and  Lemma \ref{lemma*} remains valid. The starting point for this paper is
that the Galois shuffle holds for all $W_N$ minimal models.

WZW modular data  for compact groups   at  integral level
$k\in\{1,2,3,\ldots\}$ is well-understood; their primaries $\lambda\in
\Phi$ are the level $k$ integrable highest weights for the corresponding affine Kac--Moody
algebra.  The general
method which has evolved over the years for their modular invariant classification
follows these basic steps:

\begin{itemize}

\item[\textbf{Step 1}] The vacuum row and column of a modular
invariant $M$ are heavily constrained, most significantly by
(\ref{TM}) and (\ref{parity}). In this step we solve those
constraints for $\mu=0$. In practise these constraints (usually) force $M$ to obey the condition
\begin{equation}\label{scext}
M_{0\lambda}\ne 0\ \mbox{or}\ M_{\lambda 0}\ne 0\ \Rightarrow \lambda\
\mbox{{is a simple-current}}\ .
\end{equation}

\item[\textbf{Step 2}] Find all  $M$
obeying (\ref{scext}) --- these correspond to simple-current
 extensions of the chiral algebra. For technical reasons, we first find all
$M$ obeying the stronger condition
\begin{equation}\label{perminv}
M_{0\lambda} = \delta_{0\lambda}\ \mbox{{for all}}\ \lambda
\in \Phi\ .
\end{equation}
Modular invariants $M$ satisfying (\ref{perminv}) are necessarily permutation matrices: $M_{\lambda\mu}=\delta_{\mu,
\pi\lambda}$ for some permutation $\pi$ of $\Phi$ (see e.g. Lemma 2 of \cite{Revisited} for
a proof). For that reason, these $M$ are called {\it permutation invariants}; their
importance is that multiplying by them sends modular invariants
to (usually different) modular invariants.

\item[\textbf{Step 3}] At small levels, modular invariants that do not obey (\ref{scext})
can occur.  These exceptional invariants must be classified separately.

\end{itemize}


\subsection{SU($N$) modular data and SU(3) modular invariants}\label{sA2}

In Section 3.1 we describe the $W_N(p,q)$ modular data through that of  SU$(N)$.
This subsection describes the latter, equivalently the affine algebra $A_{N-1}^{(1)}$, at integral 
levels $k$. To make some formulas cleaner, we shift the
highest weights by the Weyl vector $\rho$, use the
{\it height} $n:=k+N$ rather than the level $k$, and omit the redundant extended Dynkin label $\lambda_0$. We abbreviate  `SU$(N)$ at height $n$' with `$\SN_n$'.

The $\SN_n$ primaries can be identified with the set
\begin{equation}\label{weights}
\Phi_{N}^n := \{\lambda = (\lambda_1, \ldots,\lambda_{N-1}) \in \mathbb{Z}^{N-1} :
0 < \lambda_i,\, \lambda_1 + \cdots+\lambda_{N-1} < n\} \ ,
\end{equation}
and the vacuum (denoted 0 in Section 2.1) with  the Weyl vector $\rho = (1,\ldots,1)$.
The $T$ matrix is given by $T^{(N;n)}_{\lambda\lambda}=\alpha\exp[\pi\ii\lambda^2/n]$
for some constant $\alpha$, where $\lambda^2:=\lambda\cdot\lambda$ and
\begin{equation}\label{norm2}
\lambda\cdot \mu:=\sum_{1\le i<N}\frac{i\,(N-i)}{N}\,\lambda_i\mu_i+\sum_{1\le i<j<N}\frac{i\,(N-j)}{N}\,(\lambda_i\mu_j+\lambda_j\mu_i)\ .
\end{equation}
 The $S$ matrix entries $S^{(N;n)}_{\lambda\mu}$ are most effectively expressed as an $N\times
 N$ determinant:
 \begin{equation}\label{ssuN}
 S^{(N;n)}_{\lambda\mu}=\beta\exp[2\pi\ii\,t(\lambda)\,t(\mu)/(Nn)]\,\mathrm{det}(\exp[-2\pi\ii
 \,\lambda[i]\,\mu[j]/n])_{1\le i,j\le N}\ ,\end{equation}
where $\lambda[i]=\sum_{i\le\ell<N}(\lambda_\ell+1)$ and
 \emph{N-ality} $t(\lambda)
:= \sum_{j=1}^{N-1}j\lambda_j$.
$\beta$ is an irrelevant constant.

For any affine algebra, charge-conjugation $C$ and
simple-currents $J$ correspond to symmetries of the associated
Dynkin diagram. For SU$(N)$, they act on  $\Phi_{N}^n$ as follows:
\begin{eqnarray}
C(\lambda_1, \ldots,\lambda_{N-1})& = &(\lambda_{N-1},\ldots, \lambda_1)\ ,\label{CCSCa}\\
J(\lambda_1, \ldots,\lambda_{N-1})& =& (\lambda_0, \lambda_1,\ldots,\lambda_{N-2})\
,\label{CCSCb}
\end{eqnarray}
where $\lambda_0=n-\lambda_1-\cdots-\lambda_{N-1}$.
$C$ is order 2 and $J$ is order $N$.  They obey
(\ref{SSS'}),(\ref{TTT'}) as well as
\begin{eqnarray}
T^{(N;n)}_{J^a\lambda, J^a\mu}& =& \exp[{\pi
\ii}\,(a(N-a)n-2at(\lambda))/N]\,T^{(N;n)}_{\lambda\mu}\ ,\label{TTT}\\
S^{(N;n)}_{J^a\lambda, J^b\mu}& =& \exp[{2\pi
\ii}\,(bt(\lambda-\rho) + at(\mu-\rho) + nab)/N]\,S^{(N;n)}_{\lambda\mu}\
.\label{SSS}
\end{eqnarray}
 The quantities $Q_J$ and $h_{J0}$ in
Section 2.1 are thus $Q_{J}(\lambda)=t(\lambda-\rho)/N$ and $h_{J\rho}=(N-1)(n-N)/(2N)$.
Note that $t(\rho)\equiv_N 0$ resp. $N/2$ for $N$ odd resp. even,
   where throughout this paper `$x\equiv_my$' abbreviates `$x\equiv y$ (mod
$m$)'. Also,
\begin{equation}\label{triality}
t(J^a\lambda-\rho) \equiv_N na + t(\lambda-\rho) \ .
\end{equation}

Now specialise to $N=3$, the case of most interest to us.
The $T$ matrix for su$_{3;n}$ is
\begin{equation}\label{Tndef}
T^{(n)}_{\lambda\mu}: =T^{(3;n)}_{\lambda\mu}= \exp[2\pi \ii\frac{\lambda_1^2 +
\lambda_1\lambda_2 + \lambda_2^2 - n}{3n}]\, \delta_{\lambda\mu}\ .
\end{equation}
The denominator identity of the Lie algebra $A_2$ implies, for $1\le a<n/2$, the formula
\begin{equation}\label{denid}
S^{(n)}_{\lambda(a,a)}=S^{(n)}_{(a,a)\lambda}=\frac{8}{\sqrt{3}n}\sin\left(\frac{\pi a\lambda_1}{n}\right)\,\sin\left(\frac{\pi a\lambda_2}{n}\right)
\,\sin\left(\frac{\pi a\,(\lambda_1+\lambda_2)}{n}\right)\ ,\end{equation}
giving a convenient expression for the quantum-dimensions $\cD^{(n)}\lambda:=
S^{(n)}_{\lambda\rho}/S^{(n)}_{\rho\rho}$.

\begin{lemma}\label{lemma21} (a) For all $\lambda,\mu\in \Phi_{3}^n$, $S^{(n)}_{(2,1)\lambda}
/S^{(n)}_{\rho\lambda}=S^{(n)}_{(2,1)\mu}/S^{(n)}_{\rho\mu}$ iff $\lambda=\mu$.

\begin{itemize}

\item[(b)] Suppose $n\ne 4$. For any $\lambda\in \Phi^n_{3}$, $\lambda\not\in\langle C,J\rangle
(2,1)\cup\langle J\rangle\rho$,
$$\cD^{(n)}\lambda>\cD^{(n)}(2,1)>1$$
(when $n=4$,  $\Phi_{3}^n=\langle J\rangle\rho$).

\end{itemize}\end{lemma}

Part (a) is a special case of Proposition 3 of \cite{Symmetries} together with
$S^{(n)}_{(1,2)\nu}/S^{(n)}_{\rho\nu}=
\left(S^{(n)}_{(2,1)\nu}/S^{(n)}_{\rho\nu}\right)^*$. Part (b) is a special case of Proposition 1 of \cite{Symmetries}. (See also the proof of  Lemma \ref{lemma3|n} below.)

The Galois parity $\epsilon_\ell$ in (\ref{GalCond}) satisfies
\begin{equation}\label{su3Gal}
\eps^{(n)}_\ell(\lambda)\,\eps^{(n)}_\ell(\rho)=\left\{\matrix{
+1&\mbox{if}\ \{\ell\lambda_1\}_n+\{\ell\lambda_2\}_n<n\cr
-1&\mbox{if}\ \{\ell\lambda_1\}_n+\{\ell\lambda_2\}_n>n}\right.
\end{equation}
 for any $\ell$ coprime to $3n$, where $\{x\}_m$ is
uniquely defined by $0\le\{x\}_m<m$ and $x\equiv_m\{x\}_m$.

\begin{lemma}\label{lemma6} (a) \cite{Revisited} Suppose $\lambda\in \Phi_{3}^n$
satisfies $T_{\lambda\lambda}^{(n)\ 3}=T_{\rho\rho}^{(n)\ 3}$
and, for all $\ell$ coprime to $3n$,
$\eps_\ell^{(n)}(\lambda)=\eps^{(n)}_\ell(\rho)$. Then:

\begin{itemize}

\item[$(i)$] for $n\equiv_4 1,2,3$, $n\ne 18:$
$\lambda\in\langle J\rangle\rho;$

\item[$(ii)$] for $n\equiv_4 0$, $n\ne 12,24,60:$
$\lambda\in\langle J\rangle\rho\cup\langle J\rangle\rho''$ where $\rho''=(
\frac{n-2}{2},\frac{n-2}{2});$

\item[$(iii)$] $n=12:$ $\lambda\in\langle J\rangle\rho\cup
\langle J\rangle(3,3)\cup\langle J\rangle(5,5);$

\item[] $n=18:$ $\lambda\in\langle J\rangle\rho\cup \langle C,J\rangle(4,1);$

\item[] $n=24:$ $\lambda\in\langle J\rangle\rho\cup
\langle J\rangle(5,5)\cup\langle J\rangle(7,7) \cup\langle J\rangle(11,11);$

\item[] $n=60:$ $\lambda\in\langle J\rangle\rho\cup \langle J\rangle(11,11)
\cup\langle J\rangle(19,19)\cup\langle J\rangle(29,29)$.
\end{itemize}

\noindent(b) \cite{KR} Suppose $n$ is coprime to 6. Then
$\epsilon_\ell^{(n)}(\lambda) = \epsilon_\ell^{(n)} (\kappa)$ for
all $\ell$ coprime to $3n$, iff $\kappa\in\langle C,J\rangle\lambda$.
\end{lemma}

The modular invariants for SU$(3)$ were classified in
\cite{SU3,Revisited}, and are building blocks for those of both
$\mathrm{SU}(3)\times\mathrm{SU}(3)$ and $W_3$. The \textit{generic} su$_{3;n}$
modular invariants, existing at any height $n\ge 4$, consist of  $\bA_{n}:=I$;
charge-conjugation $\bA^*_n:=C$; the simple-current modular invariant
\begin{eqnarray}
\label{Dnonzero}
(\bD_n)_{\lambda,\mu}&=&\delta_{\mu,J^{n\,t(\lambda)}\lambda}\qquad\qquad\qquad\qquad
\qquad\qquad\mathrm{for}\ 3\nmid n\ ,\\
\label{Dzero}
&=&\left\{\matrix{\delta_{\mu\lambda}+\delta_{\mu,J\lambda}+
\delta_{\mu,J^2\lambda}&{\rm if}\ 3\mid t(\lambda)\cr 0&{\rm
otherwise}}\right.\qquad\mathrm{for}\ 3\mid n\ ,
\end{eqnarray}
given by $\bD_n:=M[J]$ in (\ref{scinv}); and the matrix product
$\bD^*_n:=C\bD_n$. The remaining modular invariants,  the \textit{exceptionals},
 in character notation (\ref{intro:part}) are
\begin{eqnarray}
\bE_8&=& |\chi_\rho + \chi_{(3,3)}|^2 + |\chi_{(1,3)}
+\chi_{(4,3)}|^2
+ |\chi_{(3,1)} + \chi_{(3,4)}|^2 \nonumber \\
&&+ |\chi_{(3,2)} + \chi_{(1,6)}|^2 + |\chi_{(4,1)} +
\chi_{(1,4)}|^2 + |\chi_{(2,3)} + \chi_{(6,1)}|^2 \,,\\
\bE_{12}&=& |\chi_\rho + \chi_{(1,10)} + \chi_{(10,1)} +
\chi_{(5,5)} + \chi_{(5,2)} + \chi_{(2,5)}|^2 + 2|\chi_{(3,3)} +
\chi_{(3,6)} +
\chi_{(6,3)}|^2 \,,\\
\bE_{12}'&=& |\chi_\rho + \chi_{(10,1)} +
\chi_{(1,10)}|^2 + |\chi_{(3,3)} + \chi_{(3,6)} + \chi_{(6,3)}|^2
+ |\chi_{(1,4)} + \chi_{(7,1)} + \chi_{(4,7)}|^2 \nonumber  \\
&&+ |\chi_{(4,1)} + \chi_{(1,7)} + \chi_{(7,4)}|^2 +
|\chi_{(5,5)}+
\chi_{(5,2)} + \chi_{(2,5)}|^2 + 2|\chi_{(4,4)}|^2\nonumber\\
&&+ (\chi_{(2,2)} + \chi_{(2,8)} + \chi_{(8,2)})\chi^*_{(4,4)} +
\chi_{(4,4)}(\chi^*_{(2,2)} + \chi^*_{(2,8)} + \chi^*_{(8,2)}) \,,\\
\bE_{24}&=& |\chi_\rho + \chi_{(5,5)} + \chi_{(7,7)} +
\chi_{(11,11)} + \chi_{(22,1)} + \chi_{(1,11)}         \nonumber  \\
&&\qquad + \chi_{(14,5)} + \chi_{(5,14)} + \chi_{(11,2)} + \chi_{(2,11)}+
\chi_{(10,7)} + \chi_{(7,10)}|^2 \nonumber\\
&&+ | \chi_{(1,7)}+\chi_{(7,1)} +  \chi_{(1,16)}+ \chi_{(16,1)} + \chi_{(7,16)}+\chi_{(16,7)}  \nonumber \\
&&\qquad +\chi_{(5,8)} +  \chi_{(8,5)} + \chi_{(5,11)} + \chi_{(11,5)} +
\chi_{(8,11)}+ \chi_{(11,8)} |^2\,,
\end{eqnarray}
at heights $n=8,12,12,24$ respectively, as well as the matrix products
$\bE_8^*:=C\bE_8$ and ${\bE'}_{12}^{*}:=C\bE'_{12}$.
Some curiousities about the SU$(3)$ modular invariants are described in \cite{bcir}.


\section{The modular invariants of $W_N(p,q)$}  \setcounter{equation}{0}

This section introduces the $W_N(p,q)$ classification problem and reduces its solution to
that of  $\mathrm{SU}(N)\times\mathrm{SU}(N)$ at height $(p,q)$ (see Theorem \ref{main}, one
of the main results of this paper). This is not at all an easy observation
and to our knowledge nothing like this has appeared in the literature before. Section 3.3 uses
this correspondence to rewrite the
 classification proof for Virasoro minimal models. The $\mathrm{SU}(3)\times\mathrm{SU}(3)$ modular invariant
 classification is given by Theorem \ref{thma2a2}
 (though its proof is deferred to Sections 4 and 5).  The
complete list of $W_3(p,q)$ modular invariants (the other main result of this paper) is Theorem \ref{th:W3ModInvs}. Previously, only
the \textit{unitary} $W_3(p,q)$ modular invariants (i.e. the special case $q=p+1$)
were classified \cite{coset}.

\subsection{The $W_N(p,q)$ modular data}\label{sw3}

Choose any integers $N\ge 2$ and  $p, q > N$. We require $p,q$ to be
coprime. The $W_N$ minimal model modular data is related  to
that of  WZW models on  SU$(N)\times\mathrm{SU}(N)$.
 We abbreviate `$\mathrm{SU}(N)\times\mathrm{SU}(N)$ at height $(p,
q)$' by `$\Ss_{p, q}$'.  A highest weight for
$\Ss_{p, q}$ is a pair $\lambda\mu:=(\lambda,
\mu)$ in $\Phi_{N}^{p,q}:=\Phi_{N}^p\times \Phi_{N}^q$.
 There are $N^2$ simple-currents for $\Ss_{p,q}$, namely $J^i K{}^j $ in obvious
 notation. The $\Ss_{p,q}$ modular data is the tensor of that of $\SN_p$ and $\SN_q$:
\begin{equation}
S^{(N;p, q)}_{\lambda\mu, \kappa\nu} = S^{(N;p)}_{\lambda\kappa}
S^{(N;q)}_{\mu\nu}\ ,\qquad    T^{(N;p, q)}_{\lambda\mu,
\lambda\mu} = T^{(N;p)}_{\lambda\lambda} T^{(N;q)}_{\mu\mu}\
.\label{SA2+data}
\end{equation}

For Theorem \ref{main} below, call  an integer $r$ \textit{$pq$-admissible} if both
 \begin{equation}\label{pqprime}
 {\ell'}:=rp-q\quad \mathrm{and}\quad \ell'':=r^2p+q\ \mathrm{are\ coprime\ to}\ 2N\,.\end{equation}
For example, when $N=2$ or 3 we can (and will) take $r=0$.
 For any $N,p,q$, there are many $pq$-admissible $r$:
  e.g. for each prime $P$ dividing $2N$, put $r_P=1$ if $P\mid pq$ and $r_P=P$
 otherwise, then $r=\prod_Pr_P$ works. Fix any $pq$-admissible $r$.
 Then (\ref{triality}) and gcd$(\ell',N)=1$ say each $JK$-orbit $\langle JK\rangle\lambda'\mu'
 =\{(J^i\lambda,K^i\mu):0\le i<N\}$  has exactly
one element $\lambda\mu$ in
  \begin{equation}
   \label{primariesWN}
\Phi_{WN}^{p,q} := \{\lambda\mu\in \Phi_{N}^{p,q} \;:\; t(\mu)\equiv_Nr\,t(\lambda)\}\,.
\end{equation}
Similarly, gcd$(\ell'',N)=1$ says each orbit $\langle J^{-r}K\rangle\lambda'\mu'$ has exactly
one element in $\Phi_{WN}^{p,q}$.

The $W_N$ minimal models are parametrised by coprime integers $p,q>N$. Their central
charge is $c=(N-1)[1-N(N+1)(p-q)^2/pq]$. $W_N(p,q)$ is unitary iff $|p-q|=1$.
 A $W_N(p,q)$ primary is a $JK$-orbit $[\lambda\mu]:=\langle JK\rangle\lambda\mu$.
We will sometimes  identify the $W_N(p,q)$ primaries with
$\Phi_{WN}^{p,q}$. The vacuum is $[\rho\rho]$. The $W_N(p,q)$ modular data is
\begin{eqnarray}
S_{[\lambda\mu] [\kappa\nu]}& =& \alpha'
\exp[-2\pi \ii\,\frac{t(\lambda)t(\nu) +t(\mu)t(\kappa)}{N}]
 \,S^{(N;p/q)}_{\lambda\kappa}S^{(N;q/p)}_{\mu\nu}\ ,
\label{W3moddataS}\\
T_{[\lambda\mu] [\lambda\mu]}& =& \beta'\exp[\pi \ii\,
\left((q\lambda- p\mu)^2\right)/(pq)]\, ,\label{W3moddataT}
\end{eqnarray}
 where $\alpha'$ and $\beta'$ are independent of $[\lambda\mu],[\kappa\nu]$, and
 $S^{(N;p/q)}$ is the matrix (\ref{ssuN}) for
$\SN_n$ formally evaluated at the fractional height
$n=p/q$.  The $N$ simple-currents of $W_N(p,q)$ are generated by
$J1$ with $Q_{J1}([\lambda\mu])= (qt(\lambda)-pt(\mu))/N +(N-1)/2$
 and $h_{[J\rho,\rho]}=(N-1)[pq+N\,(p+q)]/(2N)$.


\subsection{The Galois shuffle for $W_N$}\label{sGS}

For most $p,q$, the $W_N(p,q)$ modular data is non-unitary: $S_{[\lambda\mu] [\rho \rho]}$ can
be negative. Normally this would be bad news, as basic tools needed in modular invariant
classifications (e.g. Lemma \ref{lemma*}) break down for non-unitary modular data.
However for any $W_N(p,q)$ there is \textit{unitary} modular data $\widehat{S},\widehat{T}$
with an equivalent list of modular  invariants, obtained from $S,T$ by the \emph{Galois shuffle}
of \cite{Nonunitary}. The subtle argument  is
given in detail  in Section 6 of \cite{Nonunitary}  but only sketched below. Here we make the
crucial observation that
$\widehat{S},\widehat{T}$ can be arranged to be rescaled submatrices  of
$\Ss_{p,q}$ modular data,
permitting the association of $\Ss_{p,q}$ modular invariants with
 $W_N(p,q)$ ones:

\begin{theorem}\label{main} Fix any $pq$-admissible $r$.
Let  $M$ be any modular invariant for $W_N(p,q)$. Let $\widetilde{M}$ be the
 matrix indexed by $\Phi_{N}^{p,q}$ with entries
\begin{equation}\label{A2W3modinv}
\widetilde{M}_{J^{ar}\lambda K^{-a}\mu, J^{br}\kappa K^{-b}\nu} = \delta_{ab}
    M_{[\lambda \mu] [\kappa \nu]}\ , \end{equation}
for any $\lambda\mu,\kappa\nu\in\Phi_{WN}^{p,q}$ and any  $0\le a<N$.
 Then $\widetilde{M}$ is an $\Ss_{p,q}$
modular invariant. Conversely, an $\Ss_{p,q}$ modular invariant $\widetilde{M}$
is associated in this way to a (necessarily unique) $W_N(p,q)$ modular invariant $M$, iff
$\widetilde{M}_{J^{-r}\rho K\rho,J^{-r}\rho K\rho}=1$.
\end{theorem}

\noindent\textit{Proof of Theorem.}
 \cite{Nonunitary} proved
that for any $\ell$ coprime to $2N$ satisfying both $\ell q\equiv_p 1$  and $\ell p\equiv_q 1$,
 $[J_o\sigma_\ell \rho,\sigma_\ell \rho]$ is the $W_N(p,q)$ primary $o$  with minimal
conformal weight, where $\sigma_\ell$ is the Galois permutation in (\ref{GalCond}), and
$J_o=id.$ for $N$ odd and $J_o=id.$ or $J^{N/2}$ for $N$ even. The desired unitary modular data is
\begin{equation}
\widehat{S}_{[\lambda\mu] [\kappa\nu]} :=
\epsilon\,
\sigma_\ell(S_{[J_o\lambda,\mu][J_o\kappa,\nu]})\ ,\qquad
\widehat{T}_{[\lambda\mu][\kappa\nu]} :=
\epsilon \,(T_{[J_o\lambda,\mu][J_o\kappa,\nu]})^\ell\ ,
\label{SHat}
\end{equation}
where $\epsilon\in\{\pm 1\}$ is an irrelevant
constant. Moreover, the identity $M_{[J_o \rho,\rho][J_o\rho,\rho]}=1$ (also proved in Section 6 of
\cite{Nonunitary}) together with
Consequence 2(viii) there ensures the
bijection $M\leftrightarrow\widehat{M}$ between $W_N(p,q)$ and $\widehat{S},\widehat{T}$
modular invariants, where $\widehat{M}_{[\lambda\mu][\kappa\nu]}=M_{[\lambda
\mu][\kappa\nu]}$. Lemma \ref{lemma*}(a) now says that \begin{equation}\label{MJo}
\widehat{M}_{[J_o\lambda,\mu][J_o\kappa,\nu]}=\widehat{M}_{[\lambda\mu][\kappa\mu]}\ .
\end{equation}

To go further, fix $\ell\equiv_{2N}\ell''\ell'{}^{-2}$. It  obeys
$\ell q= 1+Ap$ and $\ell p=1+Bq$ where
\begin{equation}\label{AB}
A\equiv_{2N}\ell'{}^{-2}r\,(rq-rp+2pq)\ \mathrm{and}\ B\equiv_{2N}\ell'{}^{-2}\,(2rp+p-q)\,.
\end{equation}
Incidentally, for $N$ even, $J_o\ne id.$  here iff $r$ is even. All $pq$-admissible $r$ are even iff $pq$ is odd.

We want to show that for any $\lambda\mu, \kappa\nu \in \Phi_{WN}^{p,q}$,
\begin{eqnarray}
\widehat{S}_{[J_o\lambda,\mu][J_o\kappa,\nu]}& = &\alpha'' \,S^{(N;p,q)}_{\lambda\mu,\kappa\nu}\ ,\label{hatss}\\
\widehat{T}_{[J_o\lambda,\mu][J_o\lambda,\mu]}& =& \beta''\,T^{(N;p,q)}_{\lambda\mu,\lambda\mu}\ ,\label{hatst}
\end{eqnarray}
where $\alpha'',\beta''$ are independent of $\lambda\mu,\kappa\nu$.
 To see  (\ref{hatst}), first note from (\ref{norm2}) that for any $\lambda,\mu\in
\Phi^n_N$,
 \begin{eqnarray}
 N\lambda^2&\equiv_{2N}&(N-t(\lambda))\,t(\lambda)\ ,\\
 N\lambda\cdot\mu&\equiv_N&-t(\lambda)\,t(\mu)\ .
 \end{eqnarray}
Take any $\lambda\mu\in\Phi_{WN}^{p,q}$, and write $s:=t(\lambda)$ and $t(\mu)=:rs+s'N$. Then
 \begin{eqnarray}
 \widehat{T}_{[J_o\lambda\mu][J_o\lambda\mu]}&=&\beta'''T^{(N;p,q)}_{\lambda\mu,\lambda\mu}
 \exp[\pi\ii\,(A(N-s)s+2\ell s(rs+s'N)+B(N-rs-s'N)(rs+s'N))/N]\nonumber\\
&=& \beta''T^{(N;p,q)}_{\lambda\mu,\lambda\mu}
(-1)^{A+Br}\exp[\pi\ii\,(2\ell r-A-Br^2)s^2/N]\ .\end{eqnarray}
But both $2\ell r-A-Br^2\equiv_{2N}0$ and $A+Br\equiv_20$ follow automatically from (\ref{AB})
and (\ref{pqprime}), giving (\ref{hatst}). To see (\ref{hatss}), use (\ref{ssuN}) and (\ref{primariesWN})
 to write
 \begin{eqnarray}
 \widehat{S}_{[J_o\lambda,\mu][J_o\lambda,\mu]}&=&\alpha'''\exp[2\pi\ii\frac{-2\ell r\, t(\lambda)\,t(\kappa)}{N}+
 \frac{\ell q\,t(\lambda)\,t(\kappa)}{Np}+\frac{\ell p\,t(\mu)\,t(\nu)}{Nq}]\,
 \sigma_\ell(S'_{\lambda\kappa}{}^{(N;p/q)}S'_{\mu\nu}{}^{(N;q/p)})\nonumber\\ &=&
 \alpha''\exp[2\pi\ii\,t(\lambda)\,t(\kappa)\,(-2r\ell+A+Br^2)/N]\,S^{(N;p,q)}_{\lambda\mu,\kappa\nu}
\ ,\end{eqnarray}
where $S'$ denotes the determinant in (\ref{ssuN}). Of course $-2r\ell+A+Br^2\equiv_N0$
 so we're done.

The proof that $\widetilde{M}$ defined above commutes with $S^{(N;p,q)}$ and $T^{(N;p,q)}$
is now an easy application of (\ref{SSS}),(\ref{TTT}),(\ref{primariesWN}) and (\ref{MJo}).
 That the condition $\widetilde{M}_{J^{-r}\rho
K\rho,J^{-r}\rho K\rho}=1$ ensures a corresponding $M$ exists, follows from Lemma \ref{lemma*}(a).
 QED

\begin{corollary} The modular invariants for $W_N(p,N+1)$ are in natural one-to-one
bijection with those of SU$(N)$ at level $p-N$.\end{corollary}

Indeed, $\Phi_N^{N+1}=\langle J\rangle\rho$. Since $q=N+1$ is coprime to $N$, take $r=0$
in Theorem \ref{main}. Define $M'_{\lambda\kappa}=M_{[\lambda\rho][\kappa\rho]}$. Then
(\ref{A2W3modinv}) says $\widetilde{M}_{\lambda K^a\rho,\kappa K^b\rho}=\delta_{ab}M'_{
\lambda\kappa}$, so $\widetilde{M}$ is an $\Ss_{p,q}$ modular invariant iff $M'$ is an
$\SN_p$ one.


\subsection{A warm-up exercise: the Virasoro minimal models}

Restrict now to $N=2$.
The su$_{2;n}$ modular data is given by
\begin{equation}S_{ab}=\sqrt{{2/ n}}\,\sin\left(\pi\,{a\,
b/ n}\right)\ ,\qquad T_{aa}=\,\exp\left[{\pi\ii\, a^2/ 2n}
-{\pi\ii/ 4}\right]\ .
\end{equation}
Charge-conjugation $C$ is trivial and the vacuum is 1.
 The only nontrivial simple-current is
  $Ja=n-a$ with $Q_J(a)=(a-1)/2$ and $h_{J1}=(n-2)/4$. Recall the quantity $\{x\}_{m}$ of
Section 2.2. For any  $\ell$ coprime to $2n$,
the parity $\eps_\ell(a)$  in (\ref{GalCond}) depends on an irrelevant  contribution from $\ii\sqrt{{2/n}}$,
 as well as the sign $+1$ or $-1$,
respectively, depending on whether or not $\{\ell a\}_{2n}<n$.

The  su$_{2;n}$ modular invariants are:  the identity $I$ for all $n\ge 3$;
the simple-current invariant
\begin{eqnarray}
M[J]_{n}&=&\,\sum_{1\le a \le n-1}\,\chi_a\,\chi_{J^{a-1}a}^*\ \qquad
\qquad\qquad\qquad\qquad\qquad{\rm whenever}\ {4\mid n}\ ,\\
&=&\,|\chi_1+\chi_{J1}|^2+|\chi_3+\chi_{J3}|^2+\cdots
+2|\chi_{{n\over 2}}|^2\ \qquad {\rm whenever}\ {n\equiv_4}2\ ;\end{eqnarray}
as well as the exceptionals
\begin{eqnarray}
{\bE}_{12}^{A1}&=&\,|\chi_1+\chi_7|^2+|\chi_4+\chi_8|^2+|\chi_5+\chi_{11}|^2\
\qquad\qquad\qquad\qquad {\rm for}\ n=12\ ,\\
{\bE}_{18}^{A1}&=&\,|\chi_1+\chi_{17}|^2+|\chi_5+\chi_{13}|^2+|\chi_7+\chi_{11}|^2
\nonumber\\&&\ +\chi_9\,(\chi_3+\chi_{15})^*+(\chi_3+\chi_{15})\,\chi^*_9+|\chi_9|^2
\ \qquad\qquad\qquad {\rm for}\ n=18\ ,\\
{\bE}_{30}^{A1}&=&\,|\chi_1+\chi_{11}+\chi_{19}+\chi_{29}|^2+|\chi_7+\chi_{13}+
\chi_{17}+\chi_{23}|^2\ \qquad {\rm for}\ n=30\ .\end{eqnarray}

Cappelli-Itzykson-Zuber
\cite{CIZ} (see also \cite{CIZb,GQ,Kat}) obtained the $W_2(p,q)$ classification, together with that of su$_{2;n}$, by manifestly constructing a basis for the  commutant,
i.e. the space of all matrices commuting with $S$ and $T$. Then they imposed (\ref{intro:vac}) and
(\ref{intro:pos}).  In this way without knowing the Galois shuffle (Theorem \ref{main}) they could
still see the crucial fact that $M_{oo}=1$ and a correspondence between $W_2(p,q)$ and su$^2_{2;p,q}$.
 Even so, their proof takes several pages and involves
nontrivial number theory (e.g. that there is a prime between any $m$ and $2m$). More important,
$N=2$ behaves far simpler than $N>2$:  the generalisation of their approach to  $W_3$ (or even
SU$(3)$) has still not been found even after years of effort.

A significantly simpler proof of the su$_{2;n}$ modular invariant classification is provided
in \cite{CIZg}, using the ideas of Section 2.1. Likewise, Theorem \ref{main} permits
 a much faster proof for $W_2(p,q)$ --- indeed, much more falls out. In particular, 
Theorem 7 of \cite{SU2SU2} gives the SU$(2)\times\cdots\times\mathrm{SU}(2)$ modular
invariant classification for any height $(n_1,\ldots,n_s)$, provided only that gcd$(n_i,n_j)\le 3$ for
all $i\ne j$. Specialising to su$^2_{2;p,q}$
with $p,q$ coprime (and say $q$ odd) yields the answer:
$I_p\otimes I_q$ for all $p,q$; $M[J]_p\otimes{I}_q$ for all even $p$;
${\bE}_p^{A1}\otimes{I}_q$ for $p=12,18,30$; and  finally
\begin{eqnarray}
(M[JK]_{p,q})_{ab,cd}&=&\delta_{c,J^{a+b}a}\delta_{d,K^{a+b}b}\qquad\qquad\qquad\qquad \qquad\mathrm{whenever}\ p\equiv_4q\ ,\\
&=&\left\{\matrix{\delta_{ac}\delta_{bd}+\delta_{c,Ja}\delta_{d,Kb}&\mathrm{if}\ a
\equiv_2 b\cr 0&\mathrm{otherwise}}\right.\qquad \mathrm{whenever}\ p\equiv_4-q\ .\end{eqnarray}
This classification can also be recovered quickly by following the approach of Sections 4 and 5
(see also \cite{CIZg}).
For instance, the analogue of Lemma \ref{lemma6}(a) is: If $\epsilon_\ell(a)=\epsilon_\ell(1)$ for all
$\ell$ coprime to $2n$, then $a\in \{1,n-1\}$ unless:
$n=6$ and $a\in\{1,3,5\}$; $n=10$ and $a\in\{1,3,7,9\}$; $n=12$ and $a\in\{1,5,7,11\}$;
$n=30$ and $a\in\{1,11,19,29\}$. This is  proved
in a couple paragraphs in \cite{CIZg}.

Choosing $r=0$ in Theorem \ref{main}, we are interested in all su$^2_{2;p,q}$ modular
invariants with $M_{(1,q-1),(1,q-1)}=1$. Clearly all $M$ in factorised form $M'\otimes M''$,
i.e. $M'\otimes I_q$,
survive, but $M[JK]_{p,q}$ fails. This recovers the well-known result that the Virasoro minimal
model modular invariants correspond to pairs $M',M''$ of SU$(2)$ modular invariants.

\subsection{The SU(3)$\,\times \,$SU(3) and $W_3$
modular invariants}\label{ssPlus}

 Theorem \ref{main} says that  the $W_3(p,q)$ modular invariants, our main interest, are
 determined once the  $\su_{p,q}$ ones are.
Because of (\ref{SA2+data}), the latter  include the tensor products $M'\otimes M''$ where
 $M',M''$ are modular
invariants for su$_{3;p}$ resp. su$_{3;q}$. But tensor products won't
exhaust all of them: e.g.\ the simple-current modular invariants
\begin{eqnarray}
M[JK^{\pm 1}]_{\lambda\mu,\kappa\nu}&=& \delta_{\kappa,J^{(p+q)(t(\lambda)\pm t(\mu))}\lambda}
\delta_{\nu,K^{(p+q)(t(\mu)\pm t(\lambda))}\mu}\qquad\qquad\qquad\mathrm{for}\ 3\nmid p+q\ ,\\
&=&\left\{\matrix{\sum_{0\le i\le 2}\delta_{\kappa,J^i\lambda}\delta_{\nu,K^{\pm i}\mu}&\mathrm{if}\
t(\lambda)\equiv_3 \mp t(\mu)\cr 0&\mathrm{otherwise}}\right.\qquad \mathrm{for}\ 3\mid p+q\ , \end{eqnarray}
 aren't of factorised form for either sign. Indeed, the
modular invariant classification for $\mathrm{SU}(3)\times\mathrm{SU}(3)$ at
arbitrary height $(p, q)$ would be difficult to obtain (and
probably not terribly interesting). But much easier is when $p, q$ are coprime (we can also insist without loss of generality
that $3\nmid q$ and $p\ne 8$).
In this case the $\su_{p, q}$ classification
contains nothing unexpected:

\begin{theorem}\label{thma2a2} Let $p,q>3$ be coprime, $3\nmid q$, $p\ne 8$.
The  $\su_{p,q}$ modular invariants are:

\begin{itemize}

\item[$(i)$] the tensor products $M'\otimes M''$ for any
su$_{3;p}$, su$_{3;q}$ modular invariants $M'$ resp. $M''$;

\item[$(ii)$] the products $(C^i\otimes C^j\bD_q^l)\,M[JK{}^{\pm
1}]$ for any $i,j,l\in\{0,1\}$ ($C$ is SU(3) charge-conjugation);

\item[$(iii)$]  when $ p \equiv_3 1$ and $q=8$, the exceptionals $(C^i\otimes C^j
\bE_8)M[JK{}^{\pm 1}]$ for any $i,j\in\{0,1\}$;

\item[$(iv)$] when $p=12$ and $q\equiv_3-1$, the exceptionals $(C^i\otimes C^j)\bE_{12,q}$.
\end{itemize}
\end{theorem}

 $\bE_{12,q}$ is the matrix whose only nonzero entries are
 \begin{equation}
 (\bE_{12,q})_{\lambda\mu,\kappa\mu}=(\bE_{12,q})_{\lambda'\mu,\lambda K^a\mu}=
 (\bE_{12,q})_{\lambda K^a\mu,\lambda'\mu}=(\bE_{12,q})_{\lambda' K^a\mu,\kappa' K^b
 \mu}=1\label{exc12q}
 \end{equation}
for any $\lambda,\kappa\in[\rho]\cup[\rho'']$, $\lambda',\kappa'\in[(3,3)]$,
$ \mu\in\Phi_3^q$ with $3|t(\mu)$,  and any $a,b\in\{\pm 1\}$. See (1.1) for
 another description of $\bE_{12,5}$. This $\bE_{12,q}$
 is a modular invariant whenever $q\equiv_3-1$ ($q$ can be even). Proving modular invariance
 here reduces to verifying from (\ref{denid}) some simple identities obeyed by $S^{(12)}$, e.g.
 $S^{(12)}_{\rho(3,3)}=S^{(12)}_{(5,5)(3,3)}=S^{(12)}_{\rho\rho}+S^{(12)}_{\rho(5,5)}=-
 S^{(12)}_{(3,3)(3,3)}$.

Since $3\nmid q$,  take $r=0$ in Theorem \ref{main}, so $\Phi_{W3}^{p,q}$ consists of
all pairs $[\lambda\mu]$ with $3\mid t(\mu)$. To obtain the $W_3$ classification,
it suffices to check which
$\su_{p,q}$ modular invariants $\widetilde{M}$ have $\widetilde{M}_{\rho K\rho , \rho K \rho}=1$.
For $\widetilde{M}$ in factorised form $M'\otimes M''$, this will happen iff $M''=\bA_q$, $\bD_q^*$
or $\bE_8$. If $3\mid p+q$
then the whole $(\rho,K\rho)$-row of $M[JK^{\pm 1}]$ will vanish, whereas if $3\nmid p+q$ then
$M[JK^{\pm 1}]_{\rho K\rho,J^a\rho K^b\rho}=1$ only for $a\ne 0$. Moreover, $(\bE_{12,q})_{\rho
K\rho,J^a\rho K^b\rho}=0$ for all $a,b$. From those remarks it
is easy to confirm that any  $\widetilde{M}$ in Theorem \ref{thma2a2}(ii)-(iv)
will fail the $\widetilde{M}_{\rho K\rho , \rho K \rho}=1$ condition.

Thus any $W_3$  modular invariant $M$ corresponds via (\ref{A2W3modinv}) to an
$\widetilde{M}$ in factorised form  $M'\otimes M''$, completely analogously to
Virasoro minimal models. Explicitly, such an $M$ has entries
\begin{equation}
M_{[\lambda\mu][\kappa\nu]}=M'_{\lambda\kappa}M''_{\mu\nu}\end{equation}
where we restrict $[\lambda\mu],[\kappa\nu]$ to $\Phi_{W3}^{p,q}$, i.e. to $3\mid t(\mu)$ and
$3\mid t(\nu)$.

We can regroup these using the \emph{conjugations} $C_{W3}^{i,j}$, for $i,j,\in \:
\{0,1\}$, which act on $\Phi^{p,q}_{W3}$  by
\begin{equation}\label{W3conj}
C_{W3}^{i,j}[\lambda\mu] := [C^i\lambda, C^j\mu]\,
\end{equation}
 where $C$ is again the SU$(3)$ charge-conjugation (\ref{CCSCa}).
These $C^{i,j}_{W3}$ define $W_3(p,q)$ permutation invariants in the
usual way, and thus multiplying by them sends a modular invariant
to another.

\begin{theorem}\label{th:W3ModInvs} Assume without loss of generality that $3\nmid q$
and $p\ne 8$.
The complete list of $W_3(p,q)$ modular invariants, up to
left-multiplication by a conjugation $C_{W3}^{i,j}$, consists of  the factorised modular invariants
$\bA_p\otimes\bA_q$, $\bD_p\otimes\bA_q$, $\bE_{12}\otimes\bA_q$, $\bE'_{12}\otimes\bA_q$,
$\bE_{24}\otimes\bA_q$, $\bA_p\otimes \cE_8$ and $\bD_p\otimes\cE_8$.
\end{theorem}

As explained in Section 1, the exceptionals $(C^i\otimes C^j)\bE_{12,q}$ are not the partition
function of any RCFT.
By Theorem 3.2, all other $\su_{p,q}$ modular invariants are either tensors
of SU(3) ones, simple-current modular invariants $M[JK^{\pm 1}]$, or matrix products of
these, so they will all be realised by subfactors in the sense of \cite{BE}. (This strongly suggests
that they should also be realised in the framework of \cite{RFFS}.) Among other things, this means
they'll have a nim-rep. The corresponding $\widehat{M}$ (when it exists) for the modular
data $\widehat{S},\widehat{T}$ in (\ref{hatss}),(\ref{hatst}) will inherit this nim-rep. By
Consequence 3(i) of \cite{Nonunitary}, any of the $W_3$ modular invariants $M$ have the
same nim-reps as the corresponding $\widehat{M}$. For this reason we'd expect all $W_3$
modular invariants to give rise to an RCFT.


\section{The SU(3)$\times$SU(3) modular invariant classification}\label{s:1}
\setcounter{equation}{0}

This section proves  the $\su_{p,q}$ modular invariant classification for most $p,q$.  Sections 2.2 and 3.1 fix our
notation. For the remainder of this paper we restrict to coprime $p,q$; without loss of
generality we assume $3\nmid q$ and $p\ne 8$ (simplifying the bookkeeping).
It would have been much faster to impose from the start the $W_3$ condition $M_{\rho K\rho,
\rho K\rho}=1$ but, as the exceptionals (\ref{exc12q}) indicate, the full $\su_{p,q}$
classification is itself of some interest.


\subsection{The vacuum couplings}

Let $M$ be an $\su_{p,q}$ modular invariant, for $p,q$ as above.
When $\lambda\mu$, $\kappa\nu \in \Phi_{3}^{p,q}$ have
$M_{\lambda\mu, \kappa\nu} \neq 0$, we say that $\lambda\mu$
and  $\kappa\nu$ $M$\textit{-couple}.  The hardest step in modular invariant
classifications  is usually Step 1: to find which weights couple to
$\rho\rho$.  In the case of $\su_{p,q}$, this step
follows from Lemma 2.3(a).
   Indeed, (\ref{SA2+data}) implies that $\su_{p,q}$ Galois parities
are products $\epsilon_\ell^{(p)}(\lambda)\,\epsilon_\ell^{(q)}(\mu)$ of SU$(3)$ ones.
 By the Chinese Remainder Theorem, the parity rule  (\ref{parity}) is equivalent to
\begin{equation}\label{proofpar}
M_{\lambda\mu,\kappa\nu} \neq 0 \Rightarrow  \mbox{both}\
\epsilon^{(p)}_{\ell'}(\lambda)=\epsilon^{(p)}_{\ell'}(\kappa)\
{\rm and}\
\epsilon^{(q)}_{\ell''}(\mu)=\epsilon^{(q)}_{\ell''}(\nu)\ ,
\end{equation}
for all $\ell'$ coprime to $3p$ and all $\ell''$ coprime to $3q$. The reason is that for any such
$\ell'$, there will be an $\ell$ coprime to $3pq$ such that $\ell\equiv_p\ell'$ and $\ell\equiv_q 1$
(similarly for any such $\ell''$). Moreover, (\ref{Tndef}) and gcd$(p,q)=1$ imply from  (\ref{TM}) the
selection rule
\begin{equation}\label{prooft}
M_{\lambda\mu, \kappa\nu} \neq 0 \Rightarrow\mbox{both}\
(T^{(p)}_{\lambda\lambda})^3=(T^{(p)}_{\kappa\kappa})^3\ {\rm
and}\ (T^{(q)}_{\mu\mu})^3=(T^{(q)}_{\nu\nu})^3\ .
\end{equation}

We are now ready for the main result of this subsection. Recall the notation
$\langle J^aK^b\rangle$ for the subgroup  of simple-currents and
$\langle J^aK^b\rangle\lambda\mu$ for the corresponding orbit.

\begin{lemma}\label{Th.vacuum}

Let $p,q>3$ be coprime, $p\ne 8$,
and  $3\nmid q$.  Define $\mathcal{R}_R(M) =
\{\lambda\mu \in \Phi_{3}^{p,q} : M_{\rho\rho, \lambda\mu} \neq
0\}$ and $\mathcal{R}_L(M) = \{\lambda\mu\in \Phi_{3}^{p,q} :
M_{\lambda\mu, \rho\rho} \neq 0\}$, and $\rho'':=((p-2)/2, (p-2)/2)$, and let
$\mathcal{J}_R(M)$ (resp. $\mathcal{J}_L(M)$) consist of the simple-currents $J^aK^b$ in
$\mathcal{R}_R$ (resp. $\mathcal{R}_L$).
Then each $M_{\lambda\mu, \rho\rho},M_{\rho\rho,\lambda\mu} \in \{0,1\}$, the possibilities for
$\mathcal{J}_R=\mathcal{J}_R(M)$ and $\mathcal{J}_L=\mathcal{J}_L(M)$ are

\begin{itemize}

\item[(i)] $\mathcal{J}_R=\mathcal{J}_L=\{11\}$ \ ,

\item[(ii)] when $3\mid p:$ $\mathcal{J}_R=\mathcal{J}_L=\langle J1\rangle$ \ ,

\item[(iii)] when $3\mid p+q:$ $\mathcal{J}_R,\mathcal{J}_L\in\{\langle JK\rangle,\langle JK^2\rangle\}$ \ ,
\end{itemize}

and the possibilities for
$\mathcal{R}_R=\mathcal{R}_R(M)$ and $\mathcal{R}_L=\mathcal{R}_L(M)$ are

\begin{itemize}

\item[(a)] $\mathcal{R}_R =\mathcal{J}_R\rho\rho$, $\mathcal{R}_L = \mathcal{J}_L\rho\rho \ ,$

\item[(b)] when $q=8:$ $\mathcal{R}_R = \mathcal{J}_R\rho\rho\cup\mathcal{J}_R\rho\rho''$ \ ,
$\mathcal{R}_L = \mathcal{J}_L\rho\rho\cup\mathcal{J}_L\rho\rho''$ \ ,

\item[(c)] when $p=12:$ $\mathcal{R}_R=\mathcal{R}_L=\langle J1\rangle\rho\rho\cup\langle
J1\rangle \rho''\rho \ ,$

\item[(d)] when $p=24:$ $\mathcal{R}_R=\mathcal{R}_L=\langle J1\rangle\rho\rho\cup\langle
J1\rangle\rho'\rho\cup\langle J1\rangle\rho''\rho\cup\langle J1\rangle\rho'''\rho$ \ ,
for $\rho'=(5,5)$, $\rho'''=(7,7) \ .$

\end{itemize}\end{lemma}

\noindent\textit{Proof of Lemma.}
It suffices to look say at $\mathcal{R}_R$ and $\mathcal{J}_R$ (the equality $(MS)_{\rho\rho,\rho\rho}
=(SM)_{\rho\rho,\rho\rho}$ then pegs $\mathcal{R}_L$ and $\mathcal{J}_L$ to $\mathcal{R}_R,
\mathcal{J}_R$). If $J^aK^b\in\mathcal{J}_R$,
then $a^2p+b^2q\equiv_3 0$ by (\ref{TTT}). The only nontrivial solutions to this occur
when $3\mid p$ and $b=0$, or $3\mid p+q$ and $a=\pm b$ for either choice of sign. By Lemma
\ref{lemma*}(a), $\mathcal{J}_R$ forms a group, $\mathcal{J}_R\mathcal{R}_R=\mathcal{R}_R$,
and for all $J^aK^b\in\mathcal{J}_R$,
 \begin{equation}\label{Msc}
 M_{\rho\rho,J^a\lambda K^b\mu}=M_{\rho\rho,\lambda\mu}
 \end{equation}
  (so $M_{\rho\rho,J^a\rho K^b\rho}=1$). Among other things, this gives us (i)-(iii). It also tells
  us that when $\cR_R\subseteq \langle J\rangle\rho\times\langle K\rangle\rho$, then $\cR_R
  =\cJ_R\rho\rho$. Write $m=\|\mathcal{J}_R\|$, so
  $m=1$ or 3.

Let $\lambda\mu\in\mathcal{R}_R$, and suppose $\lambda\mu$ is not a simple-current. Then Lemma \ref{lemma6}(a)
lists the candidates for $p,q$ and $\lambda,\mu$.  We will use the condition $s_R(\kappa\nu)\ge 0$
(recall Lemma \ref{lemma*}(b)) for specific $\kappa\nu$ to prove that either  $p$ or $q$ must be
one of $8,12,24$. All $S$-matrix entries we need are
computed from (\ref{denid}), together with (\ref{SSS}) and (\ref{SSS'}).

First, suppose $p=18$. Then $4 \nmid q$ so $\mu=K^b\rho$ for some $b$. Then
$$0\le s_R((8,8)\rho)=mS^{(18)}_{\rho(8,8)}S^{(q)}_{\rho\rho}+2m\,(M_{\rho\rho,(4,1)K\rho}+
M_{\rho\rho,(4,1)K^2\rho})S^{(18)}_{(4,1)(8,8)}S^{(q)}_{\rho\rho}$$
(the factor of $m$ resp. 2 arises from (\ref{Msc}) resp. (\ref{CM})). But $S^{(18)}_{(4,1)(8,8)}/S^{(18)}_{\rho(8,8)}\approx-1.22$,
so positivity of the $S^{(n)}_{\rho\kappa}$ then forces all $C^iJ^a(4,1)\not\in\mathcal{R}_R$,
a contradiction.

Now consider $4\nmid p$. Then $\lambda\in\langle J\rangle\rho$, $4\mid q$, $\mu\in \langle K\rangle
\rho''$, and $m':=\sum_{a,b}M_{\rho\rho,J^a\rho K^b\rho''}>0$. When $q=4$ $\rho''=\rho$, so
assume $q\ge 8$. By
(\ref{Msc}), $m'$ is a multiple of $m$, hence $m'\ge m$. Then
\begin{eqnarray}\label{ineq14}
0&\le& s_R(\rho,K^q(1,2))=mS^{(p)}_{\rho\rho}S_{\rho(1,2)}^{(q)}+m'S^{(p)}_{\rho\rho}S^{(q)}_{\rho''(1,2)}\nonumber\\
&=&S^{(p)}_{\rho\rho}\frac{8}{\sqrt{3}q}\sin\left(\frac{2\pi}{q}\right)\left\{m\sin\left(\frac{\pi}{q}\right)
\sin\left(\frac{3\pi}{q}\right)-m'\cos\left(\frac{\pi}{q}\right)\cos\left(\frac{3\pi}{q}\right)\right\}\ .
\end{eqnarray}
But the right side of (\ref{ineq14}) is manifestly negative for $q\ge 12$. When $q=8$, (\ref{ineq14}) reduces to
$$m\sin(\pi/8)\sin(3\pi/8)\ge m'\sin(3\pi/8)\sin(\pi/8)\ .$$
This forces $m'=m$, i.e. $\mathcal{R}_R=\mathcal{J}_R(\rho\rho)\cup\mathcal{J}_R(\rho
\rho'')$ and $M_{\rho\rho,\rho\rho''}=1$.

Finally consider $4\mid p$. Then $q$ is odd, and the identical argument forces
$p=12,24,60$ and (using (\ref{Tndef})) $\mu=\rho$. The argument given in Section 6 of \cite{Revisited} now holds
without change (the su$_{3;q}$ component $\rho$ comes along for the ride).   QED


\subsection{The permutation invariants}\label{ss:1}

Recall from Section 2.1 that the permutation invariants are the modular invariants satisfying
$M_{\lambda\mu}=\delta_{\mu,\pi\lambda}$ for some permutation $\pi$.
 The permutation invariants for $\mathrm{SU}(N_1)\times\cdots\times\mathrm{SU}(N_s)$ at arbitrary heights
 were classified in \cite{Symmetries}. From this we read off that every permutation invariant of
  $\su_{p,q}$  is a product  $(C^i\otimes C^j)\pi_a$ for some $i,j\in\{0,1\}$ and some $\pi_a$,
where $\pi_a$ is defined as follows. To any
 $a=(a_{11}, a_{21}, a_{12}, a_{22})\in\mathbb{Z}_3^4$ obeying
\begin{equation}\label{quada}
a_{lm} + a_{ml} + pa_{l1}a_{m1} + q a_{l2}a_{m2} \equiv_3 0
\end{equation}
for all $l,m\in\{1,2\}$, define
\begin{equation}\label{sigmaa}
\pi_a(\lambda,
\mu)= (J^{a_{11}t(\lambda) + a_{21} t(\mu)} \lambda,
K^{a_{12}t(\lambda) + a_{22} t(\mu)} \mu)\ ,
\end{equation}
where $t(\lambda)$ is the triality defined in Section 2.2. The solutions to (\ref{quada}) are as follows.
When $3\mid p$, there are 6 solutions, namely $a_{11}=qa_{12}^2$ and $a_{21}=-a_{12}(1+qa_{22})$
for any $a_{22}{\not\equiv}_3-q$ and any $a_{12}$. These are all generated by $I\otimes M[K]$ and
$M[JK^{\pm 1}]$ and are included in Theorem 3.2(i)-(ii). When $3\mid p+q$, there are 4 solutions,
namely any $a_{11}{\not\equiv}_3-p$, $a_{22}{\not\equiv}_3-q$, and $a_{12}\equiv_3 a_{21}\equiv_30$. These are generated by $M[J]\otimes I$ and $I\otimes M[K]$ and are included in Theorem
3.2(i). Finally, when $3\mid p-q$, there are 8 solutions, namely any $a_{11},a_{22}\in\{0,p\}$ and
$a_{12}\equiv_3a_{21}\equiv_30$, as well as any $a_{12},a_{21}\in\{1,-1\}$ and $a_{11}\equiv_3
a_{22}\equiv_3-p$. These are generated by $M[J]\otimes I,I\otimes M[K], M[JK^{\pm 1}]$
and again are included in Theorem 3.2(i)-(ii).


\subsection{The simple-current extensions when $3\nmid p$}

The remainder of this section completes Step 2 of the $\su_{p,q}$ classification. As always, $p,q$ are coprime and $3 \nmid q$.
Let $M$ be any modular invariant satisfying both $\mathcal{R}_L(M)=\cJ_L(M)$ and
$\mathcal{R}_R(M)=\cJ_R(M)$
(these are defined in Lemma \ref{Th.vacuum}). Note that this
is automatic if neither $p$ nor $q$ is one of $8$, $12$, or $24$ (these exceptional heights are
 dealt with in Section 5). We may also assume that  $\|\mathcal{J}_R\| = \|\mathcal{J}_L\| =3$
 (otherwise $M$ is a permutation invariant).  By Lemma \ref{Th.vacuum}, there are two
 possibilities: either $3\mid p+q$ and $\cR_L,\cR_R\in\{\langle JK\rangle\rho\rho,\langle JK^2
 \rangle\rho\rho\}$ (handled in this subsection); or $3\mid p$ and $\cR_R=\cR_L=\langle J1\rangle
 \rho\rho$ (handled in the next).  The latter case  is
 more difficult as it involves \textit{fixed-points} (i.e. primaries fixed by nontrivial simple-currents
 in $\cJ_R$ or $\cJ_L$).

The following result, used in both this subsection and the next, is Lemma 3 of \cite{Revisited}.

\begin{lemma}\label{lemma**}\cite{Revisited} Let $M$ be any modular invariant for
$\su_{p,q}$ with $\mathcal{R}_L = \mathcal{J}_L\rho\rho$ and
$\mathcal{R}_R = \mathcal{J}_R\rho\rho$, and suppose that
$M_{\lambda\mu, \kappa\nu} \neq 0$.  Then $$M_{\lambda\mu,
\kappa\nu} \leq
{\|\mathcal{J}_L\|}/{\sqrt{\|\mathcal{J}_L\lambda\mu\| \
\|\mathcal{J}_R\kappa\nu\|}} \ .$$
If $\lambda\mu$ resp. $\kappa\nu$ are not fixed-points
of $\mathcal{J}_L$ resp. $\cJ_R$, then $M_{\lambda\mu, \kappa\nu} = 1$ and, in
addition, $M_{\lambda\mu, \alpha\beta} \neq 0$ iff
$\alpha\beta\in\mathcal{J}_R(\kappa\nu)$ (similarly for $M_{\alpha\beta,
\kappa\nu}\neq 0$).
\end{lemma}

Now consider $3\mid p+q$ and $\cR_R,\cR_L\in\{\langle JK\rangle\rho\rho,\langle JK^2
\rangle\rho\rho\}$. If $\cJ_R\ne\cJ_L$, hit $M$ on the left by the permutation invariant
$I\otimes\bD_q$. If
 $\cJ_R=\cJ_L=\langle JK^2\rangle$, it is also convenient  to multiply both left and right sides of
 $M$ by $I\otimes C$. Hence, without loss of generality, we can assume $\cJ_R=
 \cJ_L=\langle JK\rangle=:\cJ$. By Lemma \ref{lemma*}(b), the $\lambda\mu$-row of $M$ will
be identically zero iff $t(\lambda)\not\equiv_3-t(\mu)$ iff the $\lambda\mu$-column of $M$ is
identically zero (recall triality $t$ from Section 2.2). Let $P_{00}$ denote the set of all pairs $\lambda\mu\in \Phi^{p,q}_{3}$
such that $t(\lambda)\equiv_3t(\mu)\equiv_30$. Then any orbit $\cJ\lambda\mu$ with
$t(\lambda)\equiv_3-t(\mu)$ intersects $P_{00}$ in exactly 1 point. Because there
are no fixed-points here, Lemma \ref{lemma**} says that there is a permutation $\pi$ of
$P_{00}$ which completely determines $M$, in the sense that, for any $\lambda\mu,
\kappa\nu\in \Phi^{p,q}_{3}$, $M_{\lambda\mu,\kappa\nu}\ne 0$ iff $M_{\lambda\mu,\kappa\nu}
=1$ iff there is a $\lambda'\mu'\in P_{00}$ such that both $\lambda\mu\in\cJ\lambda'\mu'$
and $\kappa\nu\in\cJ\pi(\lambda'\mu')$.  For
$q=4$, the second component of $\pi$ is trivially the identity, so consider $q>4$.

We study  $M$ through its $\pi$. The key fact is that, for all
 $\lambda\mu,\kappa\nu\in{P}_{00}$, $SM=MS$ becomes
\begin{equation}\label{Sinv'}
S^{(p,q)}_{\lambda\mu, \kappa\nu} = S^{(p,q)}_{\lambda'\mu',
\kappa'\nu'}\  ,\end{equation}
 where we write $\pi(\lambda\mu)=\lambda'\mu'$
and $\pi(\kappa\nu)=\kappa'\nu'$. Among other things, this tells us the identity
\begin{equation}\label{qdim'}
\cD^{(p)}\lambda\,\, \cD^{(q)}\mu=\cD^{(p)}\lambda'\,\, \cD^{(q)}\mu'
\end{equation}
among quantum-dimensions $\cD^{(n)}\kappa:=S^{(n)}_{\kappa\rho}/S^{(n)}_{\rho\rho}$.
Write $\pi(\rho,K^{-q}(2,1))=\lambda'\mu'$. Suppose for contradiction that $\lambda':=(a,b)\ne \rho$. Then
by Lemma \ref{lemma21}(b), $\mu'=\rho$, and (\ref{TM}) yields
\begin{equation}\label{4.9}(a^2+b^2+ab-3)/p\equiv_14/q\ ,\end{equation}
an impossibility since $p,q$ are coprime and $q>4$. Therefore we must
have $\lambda'=\rho$, so Lemma \ref{lemma21}(b) together with (\ref{TM}) forces either $\mu'=K^{-q}(2,1)$ or
$\mu'=K^q(1,2)$. Hence we may assume $\pi(\rho K^{-q}(2,1))=\rho K^{-q}(2,1)$, multiplying on the left if necessary
by $(I\otimes C)(I\otimes\bD_q)$. Similarly, we may assume $\pi(J^{-p}(2,1)\rho)=J^{-p}(2,1)\rho$, if necessary
multiplying on the left by $(C\otimes I)(\bD_p \otimes I)$.

Now choose any $\lambda\mu\in P_{00}$ and write $\pi(\lambda\mu)=\lambda'\mu'$
as before. Comparing (\ref{Sinv'}) for $\kappa\nu=\rho\rho$ and $\kappa\nu=\rho(2,1)$ gives
\begin{equation}\label{4.10}S^{(q)}_{\mu(2,1)}/S^{(q)}_{\mu\rho}=S^{(q)}_{\mu'(2,1)}/S^{(q)}_{\mu'\rho}\ .\end{equation}
Then Lemma \ref{lemma21}(a) says $\mu=\mu'$. Using instead $\kappa\nu=(2,1)\rho$
likewise gives $\lambda=\lambda'$. Thus $\pi(\lambda\mu)=\lambda\mu$ and $M=M[JK]$.
Undoing all of the left-multiplications by permutation invariants, we recover all of Theorem
3.2(ii) when $3\mid p+q$.


\subsection{The simple-current extensions when $3\mid p$}\label{sec:fixed}

Now consider  $3\mid p$ and  $\mathcal{J}_L=\mathcal{J}_R=
\langle J1\rangle$. Here we have fixed-points, namely $\phi
\mu$ for any $\mu \in \Phi_{3}^{q}$, where $\phi := (\frac{p}{3}, \frac{p}{3})$.  By Lemma \ref{lemma*}(b), for any $\lambda\mu\in
\Phi_{3}^{p,q}$, the $\lambda\mu$-row will be nonzero iff $3\mid t(\lambda)$ (similarly for the columns).
Let $\cP_0$ be the set of $\langle J1\rangle$-orbits $[\lambda]\mu$, for all $\lambda\mu\in \Phi_{3}^{p,q}$
with $3\mid t(\lambda)$.

Lemma \ref{lemma**} says that if a nonfixed-point $[\lambda]\mu\in\cP_0$ does not couple
to a fixed-point, then $M_{[\lambda]\mu,[\lambda']\mu'}=1$ for a unique nonfixed-point
orbit $[\lambda']\mu'\in\cP_0$ (all other entries $M_{[\lambda]\mu,*}=0$). Whenever nonfixed-points
$[\lambda]\mu,[\kappa]\nu\in\cP_0$ couple to nonfixed-points $[\lambda']\mu',[\kappa']\nu'$,
then (\ref{Sinv'}) holds.

First note that every $[\rho]\mu\in\cP_0$ must couple to a nonfixed-point (otherwise (\ref{TM})
would give
$${3}/{p}\equiv_1 ({a'{}^2+a'b'+b'{}^2-a^2-ab-b^2})/{q}\ ,$$
for $\mu=(a,b),\mu'=(a',b')$, contradicting $p,q>3$ coprime). Say $M_{[\rho]\mu,[\lambda']\mu'}=1$. Then for the choice $\mu=(2,1)$, (\ref{Sinv'}),(\ref{TM}) and
Lemma \ref{lemma21}(b) force $[\lambda']=[\rho]$, by a similar argument to (\ref{4.9}). Thus
Lemma \ref{lemma21}(b) requires $\mu'=K^a(2,1)$ for some $a$, hitting $M$ on the left
if necessary by $I\otimes C$. Now (\ref{Sinv'}) for $\lambda\mu=\kappa\nu=\rho(2,1)$,
together with (\ref{SSS}), reads
$$S^{(q)}_{(2,1)(2,1)}=S^{(q)}_{K^a(2,1),K^a(2,1)}=\exp[2\pi\ii\,(-a+qa^2)/3]S^{(q)}_{(2,1)(2,1)}\ .
$$
Since $S^{(q)}_{(2,1)\lambda}\ne 0$ for any nonfixed-point $\lambda$ (this is immediate from
Lemma \ref{lemma21}(a)), we must have $a=0$ or $q$. Thus hitting $M$ if necessary by
$I\otimes\bD_q$, we can assume in fact that $M_{[\rho](2,1),[\rho](2,1)}=1$.

Now choose any $[\rho]\mu$ and write $M_{[\rho]\mu,[\lambda']\mu'}=1$. Comparing (\ref{Sinv'})
for $\kappa\nu=\rho\rho$ and $\kappa\nu=\rho(2,1)$ gives (\ref{4.10}) and hence $\mu=\mu'$;
now (\ref{qdim'}) forces $[\lambda']=[\rho]$. Thus we know $M_{[\rho]\mu,[\rho]\mu}=1$ for all $\mu$.

\begin{lemma}\label{factor} Let $M$ be any $\su_{p,q}$ modular invariant
satisfying $M_{\rho\mu,\kappa\nu}=M''_{\mu\nu} M_{\rho\rho,\kappa\rho}$ for all
$\mu,\kappa,\nu$, where $M''$ is one of $\bA_q,\bA_q^*,\bD_q$ or $\bD^*_q$. Then $M=M'\otimes
M''$ for some  su$_{3;p}$ modular invariant $M'$.\end{lemma}

\noindent\textit{Proof of Lemma.} First evaluate $MS=SM$ at $(\rho\tau,\lambda\mu)$ and
commute $S^{(q)}$ through $M''$:
$$\sum_{\kappa,\nu} M_{\rho\rho,\kappa\rho}S_{\kappa\lambda}^{(p)}S_{\tau\nu}^{(q)}M''_{\nu\mu}
=\sum_{\kappa,\nu}S^{(p)}_{\rho\kappa}S^{(q)}_{\tau\nu}M_{\kappa\nu,\lambda\mu}\ .$$
Now hit both sides with $S_{\mu'\tau}^{(q)*}$ and sum over $\tau$:
$$M''_{\mu'\mu}\sum_\kappa M_{\rho\rho,\kappa\rho}S_{\kappa\lambda}^{(p)}=
\sum_\kappa S^{(p)}_{\rho\kappa}M_{\kappa\mu',\lambda\mu}\ .$$
Hence $M_{\kappa\mu',\lambda\mu}=0$ unless $M''_{\mu'\mu}\ne 0$. For those pairs $\mu,\mu'$,
define matrices
$M'(\mu')$ by $M'(\mu')_{\lambda\kappa}=M_{\lambda\mu',\kappa\mu}/M''_{\mu',\mu}$
(for the possible $M''$ listed in Lemma \ref{factor}, $\mu$ is determined by $\mu'$ up to simple-currents $\cJ_R(M'')$,
so $M'(\mu')$ is indeed independent of $\mu$, using Lemma \ref{lemma*}(a)).

Evaluating $MS=SM$ at $(\lambda\mu,\kappa\rho)$ gives $M'(\mu)\,S^{(p)}=S^{(p)}\,M'(\rho)$,
i.e. $M'(\mu)=S^{(p)}\,M'(\rho)\,S^{(p)*}$ is independent of $\mu$ and commutes with $S^{(p)}$.
Likewise it commutes with $T^{(p)}$ and thus defines a modular invariant $M'$ of
su$_{3;p}$. QED\medskip

Hence our $M$ factorises into $M'\otimes I$ where $M'$  satisfies
$\cR_L(M')=\cR_R(M')=\langle J\rangle \rho$,
i.e. $M'=\bD_p$ or $\bE_{12}'$, and we're done. Undoing the left-multiplications yields
some of the $M$'s in Theorem 3.2(i).

(Of course the Lemma also holds with the roles of $p$ and $q$ interchanged. Lemma \ref{factor} holds more generally whenever the modular data is factorisable: $S=
S'\otimes S''$ and $T=T'\otimes T''$ --- e.g. coprimedness is not needed.)


\section{The exceptional modular invariants of SU(3)$\times$SU(3)}\label{A2A2ExcInvs} This
section handles Step 3: $M$ is a modular invariant of $\su_{p,q}$ with
$\cR_L(M)\ne\cJ_L(M)$ and $\cR_R(M)\ne \cJ_R(M)$. By Lemma \ref{Th.vacuum}, this can
only happen when one of
$p$ or $q$ is 8, 12, or 24.  Recall that we require $p\ne 8$.


\subsection{$q=8$ and $\|\mathcal{R}\|=2$}\setcounter{equation}{0}

Lemma \ref{Th.vacuum} says the only nonzero entries in the $\rho\rho$-row
and column are
 $M_{\rho\rho, \rho\rho} = M_{\rho\rho, \rho\rho''}=M_{\rho\rho'',\rho\rho}=1$.
Suppose $M_{\lambda\mu, \kappa\nu} \neq 0$.  By Lemma \ref{lemma*}(b), $\mu,\nu\in
\langle K\rangle\rho\cup\langle K\rangle\rho''\cup\langle J,C\rangle(4,1)$. From (\ref{TM})
then, we obtain that $\mu\in\langle K\rangle\rho\cup\langle K\rangle\rho''$ iff $\nu\in
\langle K\rangle\rho\cup\langle K\rangle\rho''$.

Evaluating $MS=SM$ at $(\lambda K^a\rho,\rho\rho)$ gives
\begin{equation}
\sum_{\lambda',b}M_{\lambda K^a\rho,\lambda' K^b\rho}S^{(p)}_{\lambda'\rho}S^{(8)}_{\rho\rho}
+\sum_{\lambda'',c}M_{\lambda K^a\rho,\lambda'' K^c\rho''}S^{(p)}_{\lambda''\rho}
S^{(8)}_{\rho''\rho} =S^{(p)}_{\lambda\rho}S^{(8)}_{\rho\rho} +
S^{(p)}_{\lambda\rho}S^{(8)}_{\rho\rho''} \ .
\end{equation}
But $\cD^{(8)}\rho'' = 3+2\sqrt{2}$, so
$S^{(8)}_{\rho\rho}$ and $S^{(8)}_{\rho\rho''}$ are linearly
independent over $\mathbb{Q}[e^{2\pi\ii /3p}]$, where the $S^{(p)}$ entries lie.  Therefore,
equating coefficients, we obtain
\begin{equation}\label{excep1}
\sum_{\lambda',b}M_{\lambda K^a\rho,\lambda' K^b\rho}S^{(p)}_{\lambda'\rho}=S^{(p)}_{\lambda
\rho}=\sum_{\lambda'',c}M_{\lambda K^a\rho,\lambda'' K^c\rho''}S^{(p)}_{\lambda''\rho}\ .
\end{equation}
Then $M_{\lambda K^a\rho,\lambda' K^b\rho}\ne 0$ forces $\cD^{(p)}\lambda\ge
\cD^{(p)}\lambda'$. But dually, we'd also get $\cD^{(p)}\lambda'\ge
\cD^{(p)}\lambda$. Hence  $\cD^{(p)}\lambda=\cD^{(p)}\lambda'$ (and likewise
 $\cD^{(p)}\lambda= \cD^{(p)}\lambda''$), and only one term $(\lambda',b)$ and $(\lambda'',c)$
 can appear nontrivially in the sums (\ref{excep1}).

 The Galois automorphism $\sigma_\ell$ corresponding to $\ell'=1,\ell''=-1$
 (recall Section 4.1) says $M_{\lambda\mu,\kappa\nu}=
 M_{\lambda\, C\mu,\kappa\, C\nu}$; this then means that when $a=0$, we must have
 $b=c=0$ (or the uniqueness of last paragraph would be violated).
So we have learned that for each $\lambda$, there are unique $\lambda',\lambda''$
such that $\cD^{(p)}\lambda=\cD^{(p)}\lambda'=\cD^{(p)}\lambda''$
and the only nonzero entries of the $\lambda\rho$-row of $M$ are $M_{\lambda\rho,\lambda'\rho}
=M_{\lambda\rho,\lambda''\rho''}=1$. But evaluating $MS=SM$ at $(\lambda\rho,\kappa(2,1))$ for any
$\kappa$, implies  $S^{(p)}_{\lambda'\kappa}=S^{(p)}_{\lambda''\kappa}$ $\forall\kappa$, hence
 $\lambda'=\lambda''$ for all $\lambda$.

Of course we can interchange the roles of rows and columns, and we find that the only
 nonzero entries of the $(2,1)\rho$-column are
 $M_{\,'(2,1)\rho,(2,1)\rho}=M_{\,'(2,1)\rho'',(2,1)\rho}=1$ for some $'(2,1)\in\Phi_3^n$ with
 $\cD^{(p)}('(2,1))=\cD^{(p)}(2,1)$.
 Lemma \ref{lemma21}(b) tells us $'(2,1)=C^aJ^b(2,1)$ for some $a,b$. But (\ref{TM})
 forces $b=0$ or (when $3\nmid p$) $b=p$. Therefore hitting $M$ on the left if necessary
 by the permutation invariants $C\otimes I$ and/or (for $3\nmid p$) $\bD_p\otimes I$, we
 may assume in fact that $'(2,1)=(2,1)$.

 Now evaluating $MS=SM$ at $(\lambda\rho,(2,1)\rho)$ gives $S^{(p)}_{\lambda'(2,1)}
 =S^{(p)}_{\lambda(2,1)}$, so  Lemma \ref{lemma21}(a) implies $\lambda'=\lambda$.
 We thus satisfy the hypothesis of Lemma \ref{factor} (with the roles of $M'$ and $M''$ reversed),
 and so $M=I\otimes M''$ where $M''$ is $\bE_8$ or $\bE_8^*$. These $M$ fall into Theorem
 3.2(i).


\subsection{$q=8$ for $3\nmid p$ and $\|\mathcal{R}\|=6$}

By Lemma \ref{Th.vacuum}, $p\equiv_31$ here with $\cJ_L,\cJ_R\in\{\langle JK\rangle,
\langle JK^2\rangle\}$ and $\cR_L=\cJ_L\rho\cup\cJ_L\rho''$, $\cR_R=\cJ_R\rho\cup
\cJ_R\rho''$. As in Section 4.3, we can assume without loss of generality that
 $\cJ_L=\cJ_R=\langle JK\rangle$, say.
The argument here is a  simplified version of the Section 5.1 proof
(e.g. Lemma \ref{lemma6}(b) applies to the first component). The resulting $M$ lie in
Theorem 3.2(iii).


\subsection{$q=8$ for $3 \mid p$ and $\|\mathcal{R}\| = 6$}

Lemma \ref{Th.vacuum} gives $\cR_L=\cR_R=
\{\langle J1\rangle \rho \rho, \langle J1\rangle\rho \rho''\}$.
This argument follows that of Sections 4.4 and especially 5.1, and the resulting $M$'s fall into
Theorem 3.2(i).
Section 5.1  uses Lemma  \ref{lemma21} to focus on  $(2,1)\in \Phi_{3}^p$.
This is replaced here with  $(2,2),(4,1)\in \Phi_{3}^p$:

\begin{lemma}\label{lemma3|n} Suppose $3$ divides $n>3$. Consider any $\mu,\nu\in \Phi_{3}^n$ with $t(\mu)\equiv_3t(\nu)\equiv_30$.

\begin{itemize}

\item[(a)] $\langle J\rangle\mu=\langle J\rangle\nu$ iff both $S^{(n)}_{(2,2)\mu}
/S^{(n)}_{\rho\mu}=S^{(n)}_{(2,2)\nu}/S^{(n)}_{\rho\nu}$ and
$S^{(n)}_{(4,1)\mu}
/S^{(n)}_{\rho\mu}=S^{(n)}_{(4,1)\nu}/S^{(n)}_{\rho\nu}$.

\item[(b)]  Suppose $n>12$. Provided $\mu\not\in
\langle C,J\rangle(4,1)\cup\langle J\rangle(2,2)\cup\langle J\rangle\rho$,
\begin{equation}\label{qdbnd}\cD^{(n)}\mu>\cD^{(n)}(4,1)>\cD^{(n)}(2,2)>1\end{equation}
(if $n<12$ replace $\cD^{(n)}(4,1)>\cD^{(n)}(2,2)$ here with
$\cD^{(n)}(2,2)>\cD^{(n)}(4,1)$, while if
$n=12$ replace that middle inequality with an equality).

\end{itemize}\end{lemma}

\noindent\textit{Proof of Lemma}. The starting point for proving part (a) is the identity
\begin{equation}\label{kape}
\chi_\lambda(\mu):=S^{(n)}_{\lambda\mu}/S^{(n)}_{\rho\mu}=ch_{\lambda-\rho}(-2\pi\ii\,\mu/n)\ ,
\end{equation}
where $ch_{\nu}$ is the character of the $A_2$-module with highest weight $\nu$
(see Section 13.9 of \cite{Kac} for the generalisation to all affine algebras). The $A_2$-characters
$ch_{\nu}$ for $3\mid t(\nu)$ span over $\mathbb{C}$ a subring of the character ring
of $A_2$. This subring is generated by $ch_{(1,1)},ch_{(3,0)},ch_{(0,3)}$. To see
this, it suffices (by Ch.VI, Section 3.4 of  \cite{Bour}) to prove the analogous statement
for the leading terms, namely the formal exponentials $e^\nu$, and this is immediate.

From (\ref{kape}) this means that any $\chi_{\lambda}(\mu)$ is a
polynomial in $\chi_{(2,2)}(\mu)$, $\chi_{(4,1)}(\mu)$,
and $\chi_{(1,4)}(\mu)=(\chi_{(4,1)}(\mu))^*$. So together  $\chi_{(2,2)}(\mu)=\chi_{(2,2)}(\nu)$ and
$\chi_{(4,1)}(\mu)=\chi_{(4,1)}(\nu)$ imply $\chi_{\lambda}(\mu)= \chi_{\lambda}(\nu)$ for all
$\lambda\in \Phi_{3}^n$ with $3\mid t(\lambda)$.
Hitting this with $S^{(n)}_{J^i\kappa,\lambda}$ and summing over $i=0,1,2$ and \textit{all}
$\lambda\in \Phi_{3}^n$ (the sum over $i$ projects away the weights with $3\nmid t(\lambda)$)
gives
$$\sum_i\delta_{J^i\kappa,\mu}/S^{(n)}_{\rho\mu}=
\sum_i\delta_{J^i\kappa,\nu}/S^{(n)}_{\rho\nu}$$
and hence $\langle J\rangle\mu=\langle J\rangle\nu$.

To prove part (b), first compare $\cD^{(n)}(1,4)$ and $\cD^{(n)}(2,2)$ for $n\ge 6$: we find
from (\ref{denid}) that
\begin{equation}
\frac{\partial}{\partial n}\frac{\cD^{(n)}(1,4)}{\cD^{(n)}(2,2)}=\frac{\pi}{n^2}
\frac{\cD^{(n)}(1,4)}{\cD^{(n)}(2,2)} (2c(2)-c(1)-c(5))\ ,\label{partial}\end{equation}
where $c(x)=x\cot(\pi x/n)$. For fixed $n$, $c(x)$ is concave decreasing over the interval
$0<x<n$. Hence from (\ref{partial}), $\cD^{(n)}(1,4)/\cD^{(n)}(2,2)$ monotonically increases
with $n$. We verify $\cD^{(12)}(1,4)=\cD^{(12)}(2,2)$, so $\cD^{(n)}(1,4)
<\cD^{(n)}(2,2)$ for $n<12$ and $\cD^{(n)}(1,4)>\cD^{(n)}(2,2)$ for $n>12$.

Call $(a,b)\in\mathbb{Z}^2$ \textit{admissible} if $1\le a\le b\le n-a-b$ and
$a\equiv_3b$. Then  for any
$\lambda\in \Phi_{3}^n$ with $3\mid t(\lambda)$, exactly one $C^iJ^j\lambda$ will be admissible.
Thus it suffices to consider quantum-dimensions of admissible $(a,b)$.
When $(a,b)\ne(a',b')$ are both admissible,
\begin{equation}
a\le a'\ \mathrm{and}\ b\le b'\ \Rightarrow\ \cD^{(n)}(a,b)<\cD^{(n)}(a',b')\ .
\label{qdneq}\end{equation}
To see this note that $\partial/\partial x(\sin(x)\,
\sin(x+y))=\sin(2x+y)$, so  $\cD^{(n)}(a,b)$ is an increasing function of $a$ (resp. $b$)
for fixed $b$ (resp. $a$).

Now let $\mu=(a,b)\ne (1,1),(1,4),(2,2)$ be admissible. If $a=1$ then $b\ge 7$ and $n\ge 15$, so
by (\ref{qdneq}) we get (\ref{qdbnd}). Similarly, if either $a=2$ (so $b\ge 5$),
or  both $a\ge 3$ and $b\ge 4$, we will have $n\ge 12$ and  again (\ref{qdbnd})
will follow from (\ref{qdneq}). The only remaining possibility is $\mu=(3,3)$ (for $n\ge 9$).
Then exactly as in (\ref{partial}), $\cD^{(n)}(3,3)/\cD^{(n)}(1,4)$ is an increasing function
of $n$. Moreover,  (\ref{qdneq}) says $\cD^{(n)}(3,3)>\cD^{(n)}(2,2)$, so combining
this with $\cD^{(9)}(2,2)>\cD^{(9)}(1,4)$ we obtain $\cD^{(n)}(3,3)>\cD^{(n)}(1,4)$ at $n=9$
and hence at all $n$. QED


\subsection{$p=12$ and $\|\cR\|=6$}
Here $q$ is coprime to 6 (recall Lemma \ref{lemma6}(b)), and $\cR_L=\cR_R=\langle J1\rangle\rho\rho\cup\langle J1\rangle
\rho''\rho$.  Write $[\lambda]=\langle J\rangle \lambda$. By Lemma \ref{lemma*}(b), the $\lambda\mu$-row of $M$ is nonzero iff
$[\lambda]\in\{[\rho],[\rho''],[(3,3)]\}$. Evaluating $MS=SM$ at $([\rho]\mu,[\rho]\rho)$ and
using the facts that $\cD^{(12)}\rho''=7+4\sqrt{3}$ is irrational and
$\cD^{(12)}(3,3)=1+\cD^{(12)}\rho''$, we
obtain for each $\mu$ two possibilities:  the only nonzero entries of the $[\rho]\mu$-row
are either (i) $M_{[\rho]\mu,[\rho]\mu^1}=M_{[\rho]\mu,[\rho'']\mu^2}=1$ for $\mu^1,\mu^2\in
\langle C,K\rangle\mu$, or (ii) $M_{[\rho]\mu,[(3,3)]\mu^3}=1$ for some $\mu^3\in\langle C,K
\rangle \mu$. Call $\mu$ `type (i)' or `type (ii)' resp. In type (i),  $\mu^1=\mu^2$ follows from the
$([\rho]\mu,[2,2]\nu)$-entries of $MS=SM$, for all $\nu$.

Suppose first that $M_{[\rho]\,{}^1(2,1),[\rho](2,1)}=1$ for some $^1(2,1)=C^aK^b(2,1)$. Hitting
$M$ with some $I\otimes C^i \bD_q^j$, we can require $^1(2,1)=(2,1)$.
$MS=SM$ at $([\rho]\mu,[\rho](2,1))$ gives either $\mu^1=\mu$ or $\mu^3=\mu$, whichever
is appropriate, but the latter violates (\ref{TM}). Therefore  Lemma \ref{factor} implies $M=\bE_{12}\otimes I$ lies in Theorem 3.2(i).

It thus suffices to suppose  $M_{[(3,3)]\,{}^3(2,1),[\rho](2,1)}=1$ for some $^3(2,1)=
 C^aJ^b(2,1)$. We can force $a=0$, but (\ref{TM}) says
 $q\equiv_3-1$ and $b=1$. For any $\mu$, $MS=SM$ at $([\rho]\mu,[\rho](2,1))$ yields
 $S^{(q)}_{\mu^j(2,1)}=S^{(q)}_{\mu,K(2,1)}$, i.e. $\mu^j=J^{t(\mu)}\mu$, for $j=1$ or 3, whichever
 is appropriate. But then (\ref{TM}) forces $\mu$ to be of type (i) if $3\mid t(\mu)$ and
 type (ii) otherwise. When $\mu$ is type (i), the Galois automorphism $\sigma_\ell$ (recall
 (\ref{GalCondDer})) with $\ell'=5,\ell''=1$ (recall Section 4.1) now also gives
  $M_{[\rho'']\mu,[\rho'']\mu}=
 M_{[\rho'']\mu,[\rho]\mu}=1$.

Finally, evaluate $MS=SM$ at $([(3,3)]\mu,[\rho]\nu)$ for any $\mu,\nu$: this gives
\begin{eqnarray}&\sum_{\mu'}M_{[(3,3)]\mu,[\rho]\mu'}S^{(12)}_{\rho\rho}S^{(q)}_{\mu'\nu}+
\sum_{\mu''}M_{[(3,3)]\mu,[\rho'']\mu''}S^{(12)}_{\rho''\rho}S^{(q)}_{\mu''\nu}+
\sum_{\mu'''}M_{[(3,3)]\mu,[(3,3)]\mu'''}S^{(12)}_{(3,3)\rho}S^{(q)}_{\mu'''\nu})&\nonumber\\ &=
S^{(12)}_{(3,3)\rho}(S^{(q)}_{K^{t(\mu)+1}\mu,\nu}+S^{(q)}_{K^{t(\mu)-1}\mu,\nu}\ .&\end{eqnarray}
Now multiply by $S^{(q)*}_{\nu\gamma}$ and sum over all $\nu$: writing $x=7+4\sqrt{3}$,
this becomes
\begin{eqnarray}&\sum_{\mu'}M_{[(3,3)]\mu,[\rho]\mu'}\delta_{\mu'\gamma}+
x\sum_{\mu''}M_{[(3,3)]\mu,[\rho'']\mu''}\delta_{\mu''\gamma}+
(1+x)\sum_{\mu'''}M_{[(3,3)]\mu,[(3,3)]\mu'''}\delta_{\mu'''\gamma}&\nonumber\\&=
(1+x)(\delta_{K^{t(\mu)+1}\mu,\gamma}+\delta_{K^{t(\mu)-1}\mu,\gamma})\ ,&\end{eqnarray}
valid for all $\gamma\in\Phi_3^q$. Using (\ref{TM}), this is enough to see $M$ is the exceptional $\bE_{12,q}$.

\subsection{$p=24$ and $\|\cR\|=12$}
Again $q$ is coprime to 6 so Lemma \ref{lemma6}(b) applies. Recall $[\lambda]=\langle J\rangle \lambda$. From Lemma \ref{Th.vacuum}
we have $\cR_L=\cR_R=\cup_{i=1}^4[\rho^i]\rho$ where for $i=1,2,3,4$ we write $\rho^i=\rho,
(5,5),(7,7),(11,11)$ respectively.

The 4 numbers $S^{(24)}_{\rho,\rho^i}$ are linearly independent
 over $\mathbb{Q}$. This is easy to see using the Galois automorphisms $\sigma_5,\sigma_7$:
   those $\sigma$'s permute the numbers $z_i:=\ii\sqrt{3}S^{(24)}_{\rho\rho^i}$
 via the permutations $(12)(34),(13)(24)$ respectively, so any linear relation of the
 form $\sum_ir_iz_i$, $r_i\in\mathbb{Q}$, is quickly seen to be trivial. This means $MS=SM$
 at $([\rho]\mu,[\rho]\rho)$, for any $\mu$, forces there to be weights $\mu^i\in\langle C,K\rangle
 \mu$ such that the only nonzero entries of the $[\rho]\mu$-row
 are $M_{[\rho]\mu,[\rho^i]\mu^i}=1$.
 That $\mu^1=\mu^2=\mu^3=\mu^4$, follows from $MS=SM$ at $(\rho\mu,(2,2)\nu)$:
 hitting it with $S^{(q)*}_{\nu\nu'}$ and summing over $\nu$ yields
 $$S^{(24)}_{\rho^1(2,2)}(\delta_{\mu^1\nu'}-\delta_{\mu^4\nu'})=S^{(24)}_{\rho^2(2,2)}(
 \delta_{\mu^3\nu'}-\delta_{\mu^2\nu'})\ .$$

As usual (hitting $M$ if necessary by $(1\otimes C^i\bD_q^j$) we can assume the only
nonzero entries of the $[\rho](2,1)$-column are $M_{[\rho^i](2,1),[\rho](2,1)}=1$.
Evaluating $MS=SM$ at $([\rho]\mu,[\rho](2,1))$ and using Lemma \ref{lemma21}(a)
gives $\mu^1=\mu$ for all $\mu$. Hence by Lemma \ref{factor},
$M=\bE_{24}\otimes I$ lies in Theorem 3.2(i).

\vspace{0.2cm}\addtolength{\baselineskip}{-2pt}
\begin{small}
\noindent{\it Acknowledgement.}
EB thanks the Department of Mathematical and
Statistical Sciences at the University of Alberta for support during this
research.
TG thanks the Institut f\"ur Mathematik of Universit\"at W\"urzburg for generous hospitality while researching
this paper. This research was supported in part by the National Sciences and Engineering
Research Council of Canada and the Deutscher Akademischer Austausch Dienst of
Germany.

\end{small}

\end{document}